\documentclass[twoside,slac_one]{revtex4}
\usepackage{graphicx}
\usepackage{fancyhdr}
\usepackage{amsmath} 
\usepackage{bm}
\usepackage{amsxtra}
\usepackage{amssymb}
\usepackage{amsthm}
\usepackage{latexsym}
\usepackage{lscape}

\begin{document}

\title{Beyond the Desert: Tevatron and LHC Results on
  Searches for Physics Beyond the Standard Model}

%

\author{G. Redlinger}
\affiliation{Physics Department, Brookhaven National Laboratory,
  Upton, NY, USA}

\begin{abstract}
  This is a brief and limited review of searches for physics beyond
  the Standard Model from the ATLAS,CDF,CMS and D0 experiments, as of
  the end of July 2011.  Priority is given to the most recent results
  and to those with the largest integrated luminosity analyzed.
\end{abstract}

\maketitle

\thispagestyle{fancy}


\section{Introduction}
As the searches for physics beyond the Standard Model (SM) at the LHC
get fully underway 
this year, the desert is on the minds of many people.  The
desert has special meaning in particle physics of course, representing perhaps
our greatest hope and our greatest fear.  The hope is that we discover
new physics at the LHC, and that this new physics
holds in the desert scenario all the way up to the GUT scale.
The fear is that we discover
nothing, a fear echoed by Bob Park in his widely read ``What's
New'' column where, on the occasion of the
first major release of results from the LHC
experiments, he reminisces about the look on Carl Sagan's face
32 years ago when the first Mars lander sent back desolate images of the
Martian desert \cite{Park}.

Given the large number of search results from the Tevatron and LHC
experiments, a comprehensive review is beyond the scope of this
document, and some selection/organizing principle is needed.
There is a temptation to organize by theoretical concepts, but this
violates the spirit of Beyond SM (BSM) searches which are typically
signature based and not tied to specific models.\footnote{Although
  most analyses do include interpretations in one or more models.}
However, a full review of all
possible experimental signatures is also beyond the scope of this
note.

The note is therefore organized as follows.  For the Tevatron
searches, after a brief listing of the most recent results from a
rich program, the focus
will instead be on two of the ``anomalies'' that have caught the attention
of the community recently.  This is followed by a
review of the most recent LHC results up to the end of July,
where priority is given to the most recent results with the
largest integrated luminosity analyzed.   A full listing of
all analyses can be found at \cite{all_results}.
The reader should also
consult \cite{Stone} and \cite{Tollefson} from this conference
for other search results involving
heavy flavor (i.e. bottom and charm) and top quarks, respectively.

\section{Results from the Tevatron}

At the Tevatron, about 11.5 $\rm{fb}^{-1}$ have been delivered to each
experiment and a little over 10 $\rm{fb}^{-1}$ recorded.  Recent
search results from CDF and D0 involve 5-6 $\rm{fb}^{-1}$ of data,
although some go as high as 9 $\rm{fb}^{-1}$.
Tables \ref{D0_summary} and \ref{CDF_summary} summarize some of the most recent BSM
searches from D0 and  CDF.  Among the noteworthy new results are: {\it
  i)} the
diphoton resonance search by CDF \cite{CDF_gg_resonance} which,
combined with results from dilepton resonances,  set a
new limit on Randall-Sundrum gravitons, {\it ii)} a long-lived charged particle
search from D0 \cite{D0_LLP} which set new limits on higgsino-like and
gaugino-like charginos, and {\it iii)} the anomaly in the D0 like-sign dimuon
charge asymmetry
measurement  
\cite{D0_dimuon} where a $3.9\sigma$ deviation from SM
expectations was seen.

Other sightings at the Tevatron that have caught the attention of the
community recently are the peak in the dijet mass spectrum in W+2jets
events and the forward-backward asymmetry in $t\overline{t}$
production.

\begin{table}[ht]
\begin{center}
\caption{Incomplete summary of some of the most recent BSM
  searches from D0.}
\begin{tabular}{|l|c|l|l|}
\hline             & $\int\mathcal{L}dt$ & &\\
\textbf{Signature} & \textbf{($\rm{fb}^{-1}$)} & \textbf{Ref.} & \textbf{Comment}
\\
\hline
long-lived slow particle & 5.2 & \cite{D0_LLP} & M $>$ 230 (251)
GeV for higgsino(gaugino)-like chargino \\
dimuon asymmetry & 9 & \cite{D0_dimuon} & $3.9\sigma$ deviation from SM\\
W' $\rightarrow$ tb & 2.3 & \cite{D0_W'} & M(W') $>$ 863 GeV \\
WW/WZ resonance     & 5.4 & \cite{D0_WW} & M(RS
graviton) $>$ 300-754 GeV\\
$e\nu jj$ resonance  & 5.4 & \cite{D0_enujj} & M(LQ(1)) $>$ 326
GeV ($BR=0.5$) \\
$t' \rightarrow qW$ & 5.3  & \cite{D0_t'} & M($t'$) $>$ 285
GeV ($2.5\sigma$ excess in $\mu$ channel) \\
$\gamma\gamma + E_{T}^{miss}$ & 6.3 & \cite{D0_ggMET} &
$\Lambda > 124$ TeV (minimal GMSB) \\
$\tilde{t} \rightarrow b l \tilde{\nu} (l=e,\mu)$ & 5.4 &
\cite{D0_stop} & M($\tilde{t}$) $>$ 210 GeV for M($\tilde\nu$) $<$ 110 GeV \\
\hline
\end{tabular}
\label{D0_summary}
\end{center}
\end{table}

\begin{table}[ht]
\begin{center}
\caption{Incomplete summary of some of the most recent BSM
  searches from CDF.}
\begin{tabular}{|l|c|l|l|}
\hline             & $\int\mathcal{L}dt$ & &\\
\textbf{Signature} & \textbf{($\rm{fb}^{-1}$)} & \textbf{Ref.} & \textbf{Comment}
\\
\hline
$\gamma\gamma$ resonance & 5.7 & \cite{CDF_gg} & M(RS graviton)
$>$ 1111 GeV, ($ee$,$\mu\mu$,$\gamma\gamma$ combined,
$k/\overline{M}_{pl} = 0.1$) \\
ZZ resonance ($ll\nu\nu$,$llll$,$lljj$) & 6 & \cite{CDF_ZZ} &
M(RS graviton) $>$ $\sim$ 600 GeV \\
$l + \gamma$ + $E_{T}^{miss}$ + bjet (also
$t\overline{t}+\gamma$) & 6 & \cite{CDF_ttg} &
$\sigma(t\overline{t}+\gamma) = 0.18 \pm 0.07$ pb\\
$jjj$ resonance & 3.2 & \cite{CDF_jjj} & cross section
limits vs mass for RPV gluino \\
$t' \rightarrow b+W$ & 5.6 & \cite{CDF_t'bW} & M($t'$) $>$ 358
GeV\\
$b' \rightarrow t+W$ & 4.8 & \cite{CDF_b'tW} & M($b'$) $>$ 372 GeV\\
$t' \rightarrow t+ E_{T}^{miss}$ & 5.7 & \cite{CDF_ttMET} & M($t'$)
$>$ 400 GeV for M(inv) $<$ 70 GeV\\
$t\overline{t}$ resonance & 4.8 & \cite{CDF_ttres} & M($Z'$) $>$ 900
GeV\\
$ZZ + E_{T}^{miss}$ & 4 & \cite{CDF_ZZMET} & $\sigma > 300$ fb\\
same-sign dilepton & 6.1 & \cite{CDF_SS} & consistent with SM \\
$bbb$ resonance & 2.6 & \cite{CDF_bbb} & $\sigma$ limits as function
of mass\\
\hline
\end{tabular}
\label{CDF_summary}
\end{center}
\end{table}

\subsection{Dijet mass spectrum in W+2jets events}

\begin{figure}
\begin{minipage}{0.3\linewidth}
\begin{center}
\includegraphics[width=0.95\linewidth]{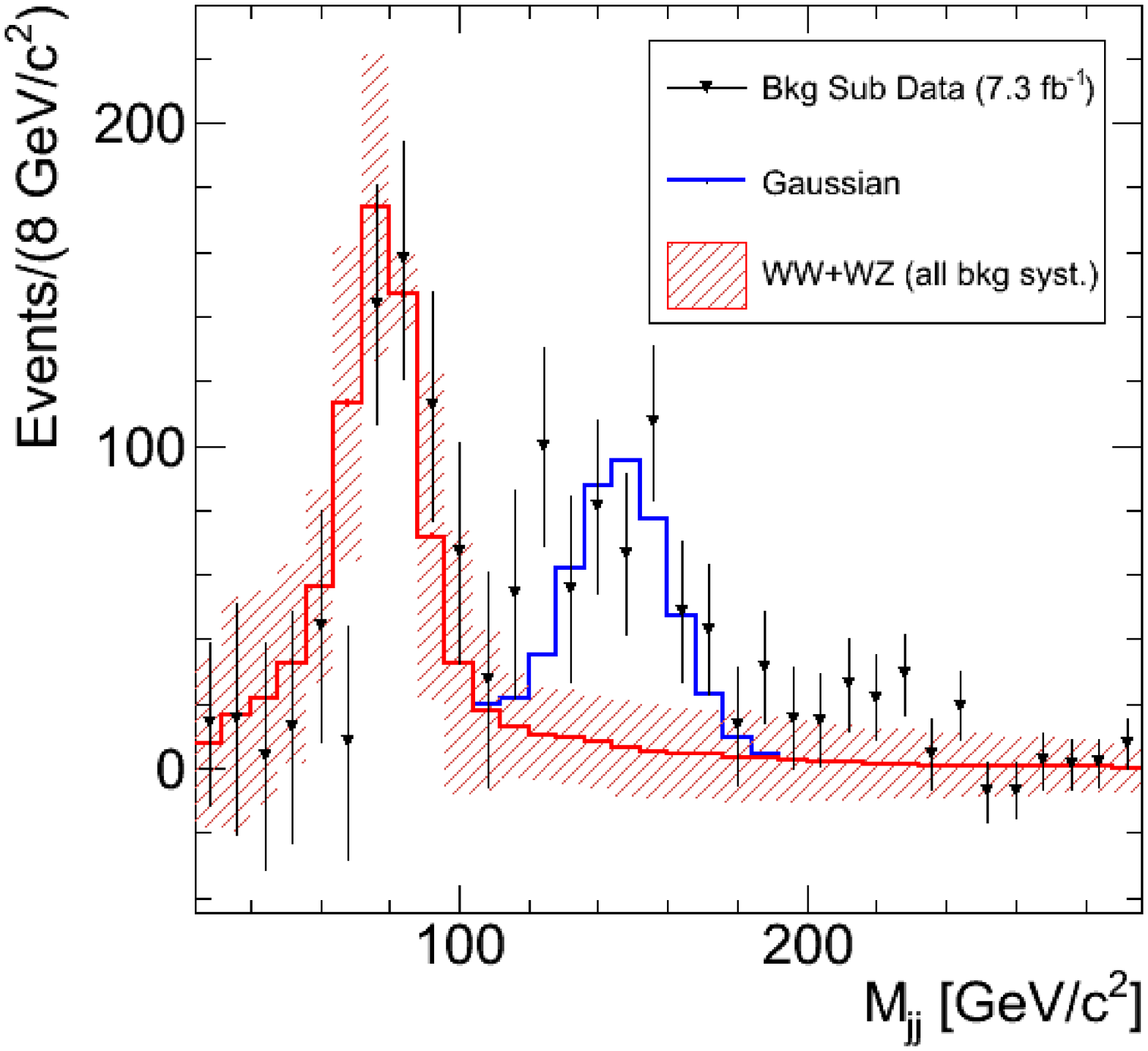}
\end{center}
\end{minipage}
\begin{minipage}{0.38\linewidth}
\begin{center}
\includegraphics[width=0.95\linewidth]{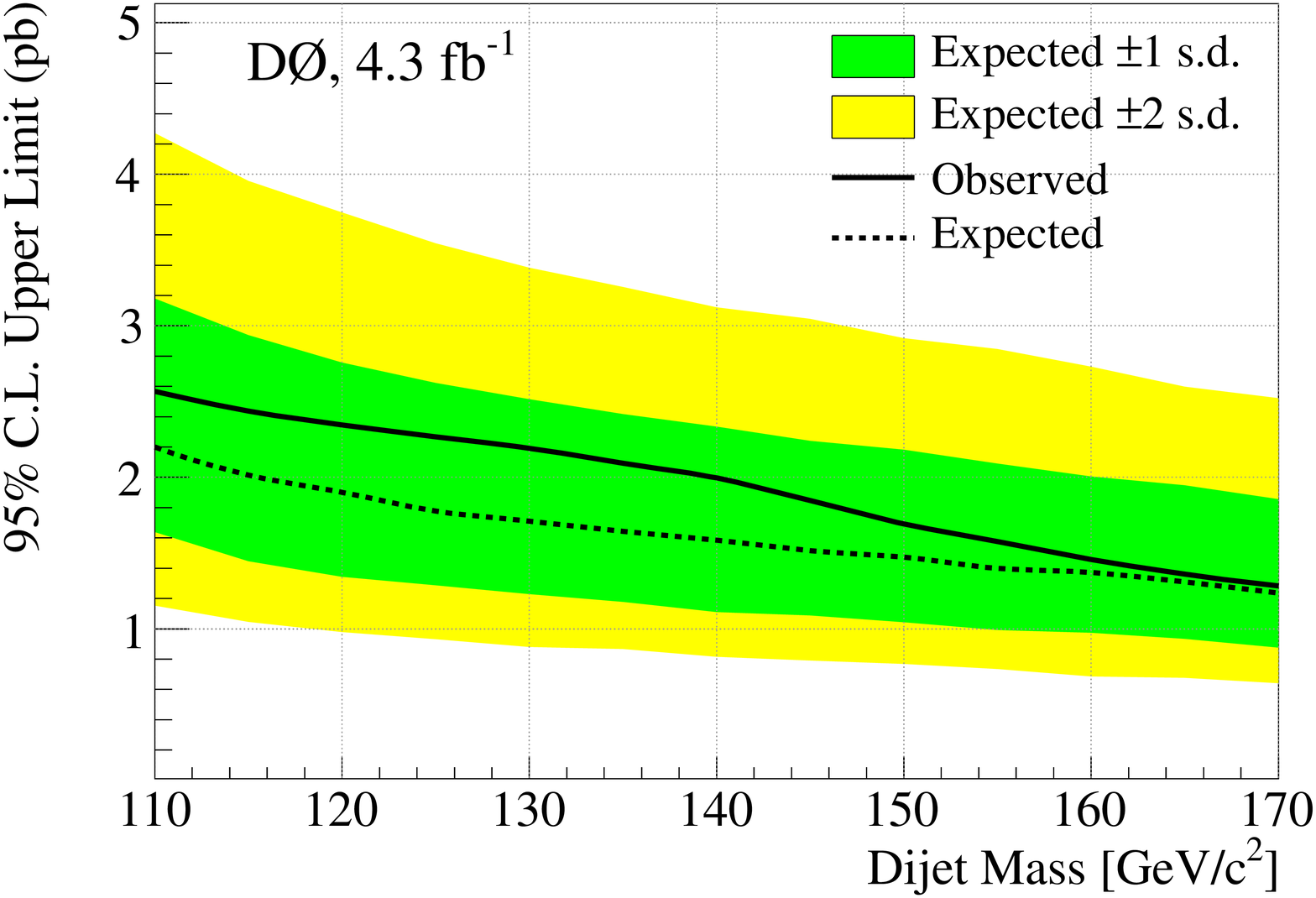}
\end{center}
\end{minipage}
\begin{minipage}{0.3\linewidth}
\begin{center}
\includegraphics[width=0.95\linewidth]{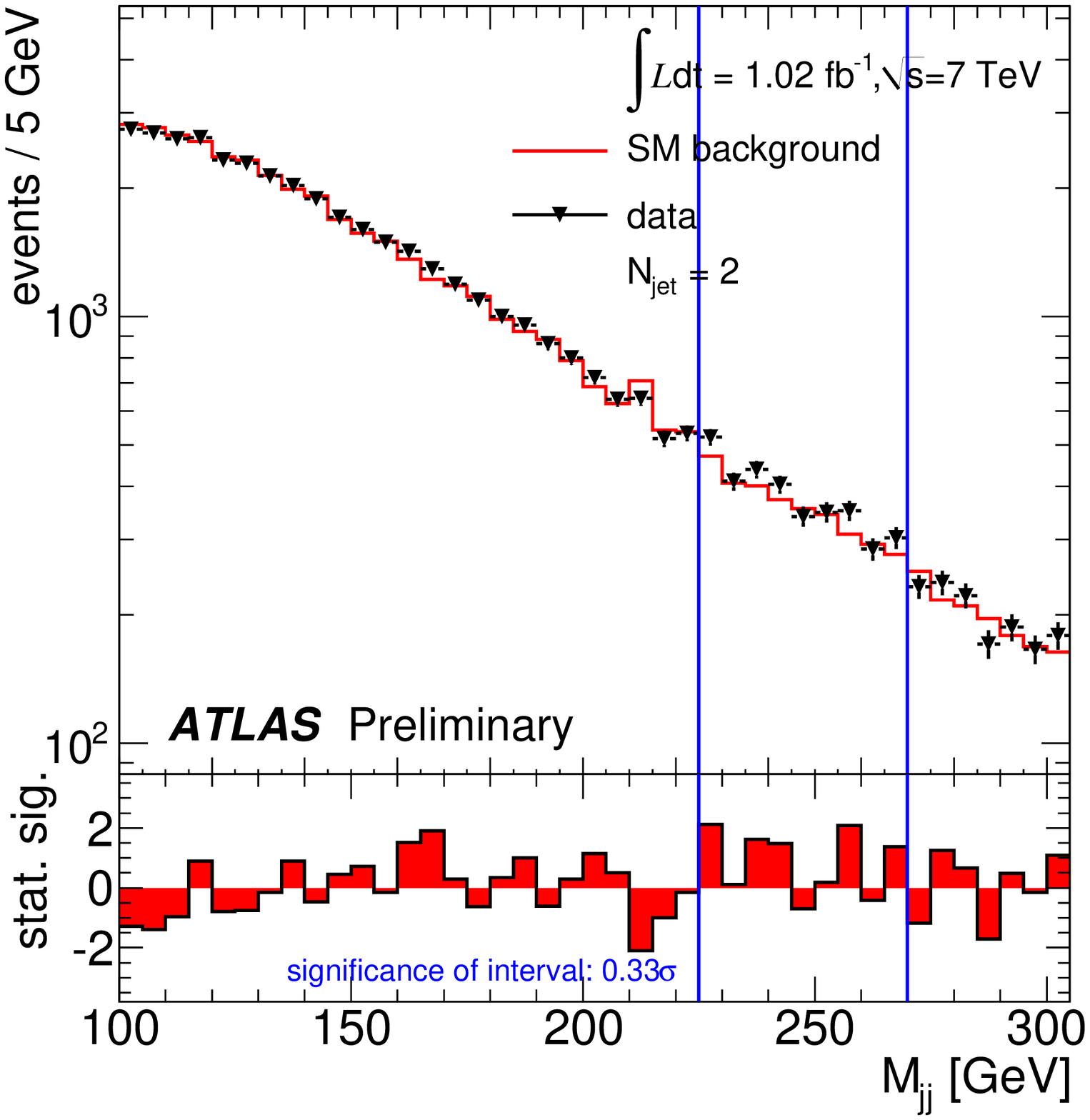}
\end{center}
\end{minipage}
\caption{Left: Dijet mass spectrum in W+2jet events from CDF
  \cite{CDF_wjets_update}. Center: D0 limits on the cross section for
  an excess as a function of the dijet mass \cite{D0_wjets}. Right:
  Dijet mass spectrum in W+2jet events from ATLAS \cite{ATLAS_wjets}.}
\label{fig:CDF_wjets}
\end{figure}

CDF reported earlier this year \cite{CDF_wjets_orig} on
an excess in the dijet mass spectrum in the 120-160 GeV range based on
an analysis of 4.3 $\rm{fb}^{-1}$ of data.  The analysis has been
recently extended \cite{CDF_wjets_update} to 7.3 $\rm{fb}^{-1}$. Using
the same analysis cuts and strategy, the excess in the dijet mass
distribution, shown in Fig.
\ref{fig:CDF_wjets}(left),  increases from
$3.2\sigma$ to $4.1\sigma$.  The estimated cross section of the excess
is $3.0 \pm 0.7$ pb, assuming the acceptance follows that of W+Higgs
production.

D0 \cite{D0_wjets} does not confirm the excess, following the same
  analysis methods on 4.3 $\rm{fb}^{-1}$ of data.  Accordingly, 
  limits are set on the production cross section, shown in
  Fig. \ref{fig:CDF_wjets}(middle), excluding the original
  crude estimate by CDF of the cross section of 4 $\rm{pb}^{-1}$ by a
  significant margin.  However, the situation is a little bit more ambiguous
  now, given the most recent estimate of the cross section from CDF.
  There had also been some worry that the Monte Carlo tuning done by
  D0 would have erased the effect, but it has been shown that these
  correction factors have very little effect.
  ATLAS \cite{ATLAS_wjets} has also made a search  in their W+2jets
data. They too come up empty; the ATLAS mass spectrum in W+2jets
events is shown
in Fig. \ref{fig:CDF_wjets}(right).

\subsection{$t\overline{t}$ forward-backward asymmetry}

\begin{figure}
\begin{minipage}{0.32\linewidth}
\begin{center}
\includegraphics[width=0.95\linewidth]{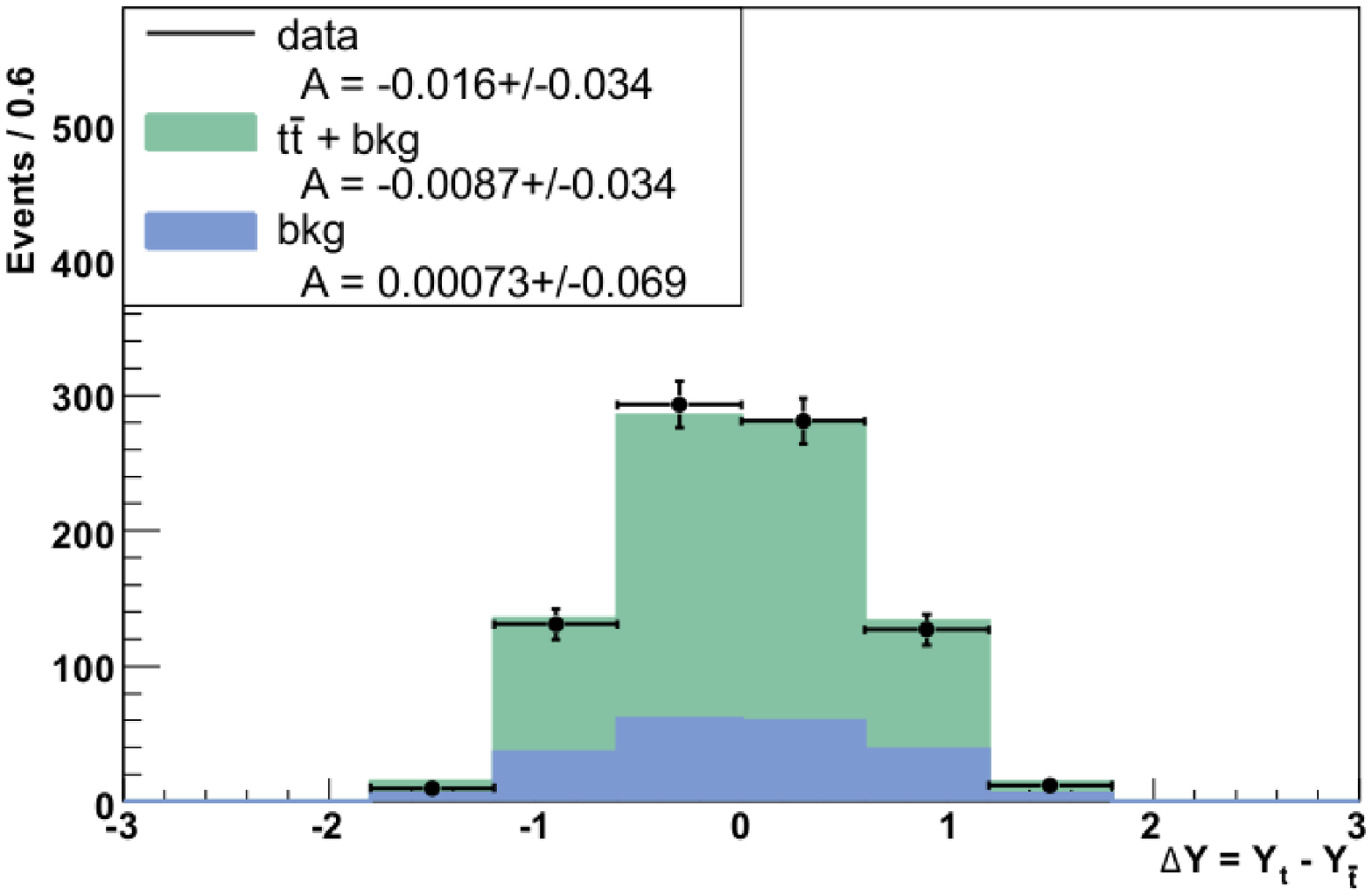}
\end{center}
\end{minipage}
\begin{minipage}{0.32\linewidth}
\begin{center}
\includegraphics[width=0.95\linewidth]{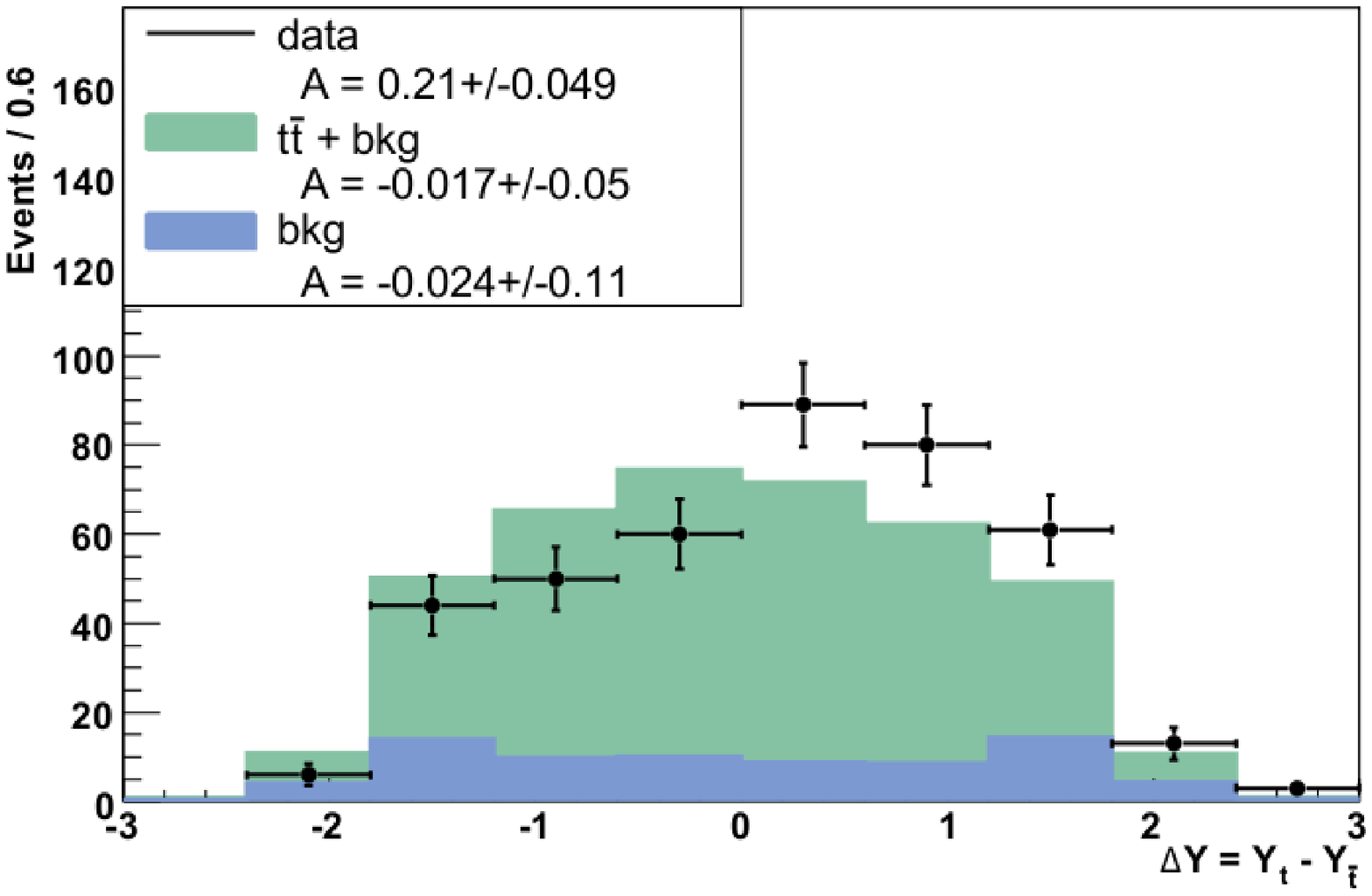}
\end{center}
\end{minipage}
\begin{minipage}{0.34\linewidth}
\begin{center}
\includegraphics[width=0.95\linewidth]{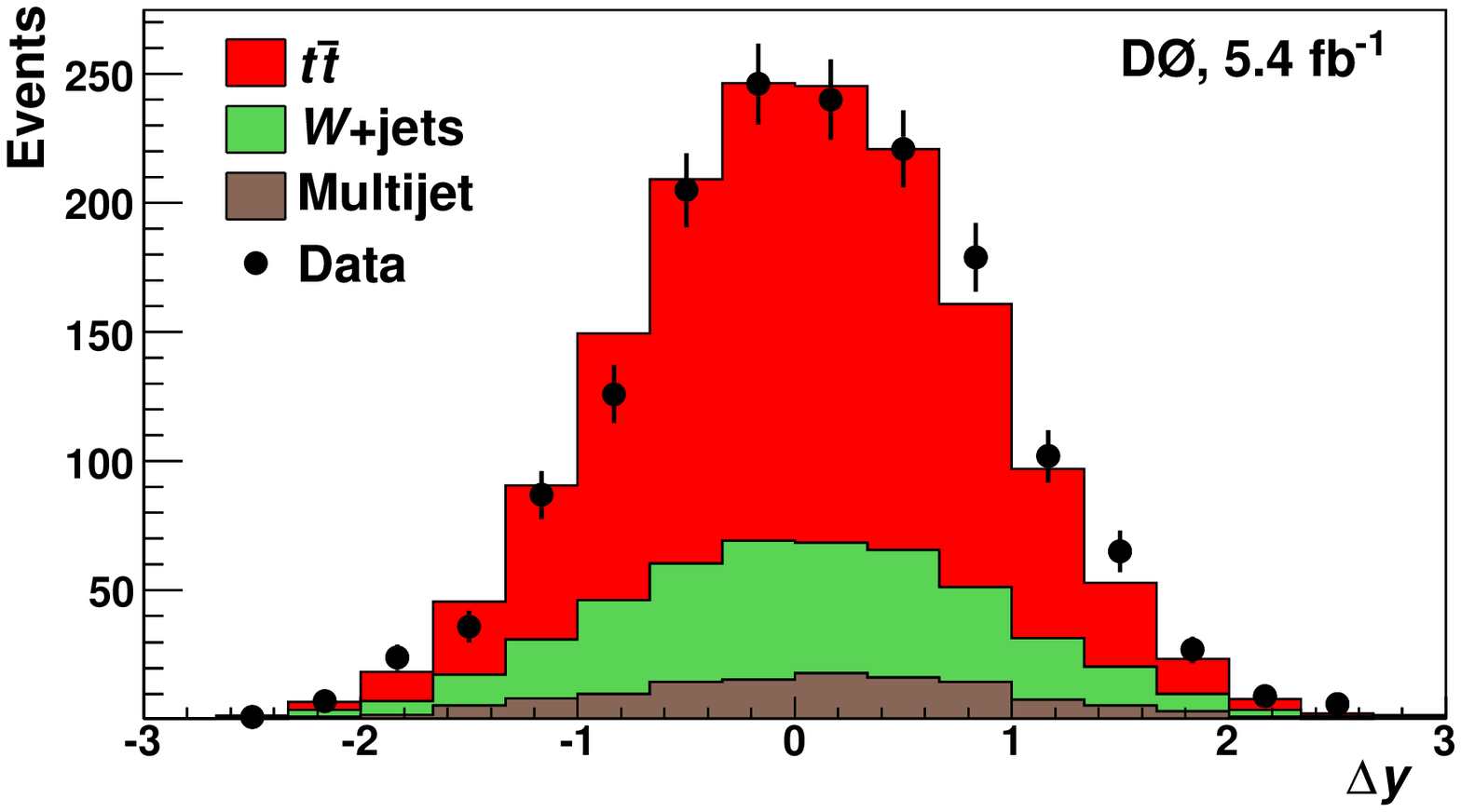}
\end{center}
\end{minipage}
\caption{Left and Center: Distributions of $\Delta y$ in
  $t\overline{t}$ events from CDF
  \cite{CDF_afb_ljets} in the lepton+jets channel.
  Left: for  $M(t\overline{t}) <
  450$ GeV. Center: for $M(t\overline{t})
  > 450$ GeV. Right: Distribution of $\Delta y$ from D0
  \cite{D0_afb_ljets} in the lepton+jets channel.}
\label{fig:tt_asym}
\end{figure}

Turning to $t\overline{t}$ production, CDF \cite{CDF_afb_ljets} has
compared the production rate in the forward and backward hemispheres
with 5.3 $\rm{fb}^{-1}$ of data.  A small asymmetry is expected in the
SM from interference between leading-order and next-to-leading order
diagramss.  
CDF first observed an asymmetry in the lepton+jets channel;
the effect is particularly enhanced
at large values of the mass of the $t\overline{t}$ pair and also at
large values of the rapidity difference between $t$ and
$\overline{t}$.
Fig. \ref{fig:tt_asym} (left) shows the distribution of $\Delta y$ for
$M(t\overline{t}) < 450$ GeV while Fig. \ref{fig:tt_asym} (center)
shows the distribution for $M(t\overline{t}) > 450$ GeV.
The asymmetry, corrected back to parton level and evaluated in the
$t\overline{t}$ rest frame is $A^{t\overline{t}} = 0.475 \pm 0.114$
for $M(t\overline{t}) > 450$ GeV,
to be compared with the prediction using MCFM \cite{MCFM},
$A^{t\overline{t}} = 0.088 \pm 0.013$, amounting to a $3.4\sigma$
discrepancy.

More recently, CDF has studied the
dileptonic $t\overline{t}$ channel \cite{CDF_afb_ll} and found a
$2.3\sigma$ deviation, $A^{t\overline{t}} = 0.42 \pm 0.15 (\rm{stat})
\pm 0.05 (\rm{sys})$, again for $M(t\overline{t}) > 450$ GeV.
D0 has analyzed the lepton+jets channel with 5.4 $\rm{fb}^{-1}$ of
data \cite{D0_afb_ljets}.  They measure the asymmetry in two ways,
based on $\Delta y$ and also a simpler analysis based on the lepton
direction.   They also observe an asymmetry significantly
above SM expectations,
but no particular enhancement
at large $\Delta y$ or at large $M(t\overline{t})$.  Whether this is a
sign of New Physics or points to a deficiency in higher-order QCD
calculations remains an open question.  As pointed out by D0, some
Monte Carlo generators predict a dependence of the asymmetry on the
$p_{T}$ of the $t\overline{t}$ system.  They find that
their measurement of $p_{T}(t\overline{t})$ comes out softer than
Monte Carlo predictions, which would lead to a larger asymmetry
predicted in the SM.

Due to the fact that the LHC is a proton-proton machine, there is no
forward-backward asymmetry.  However, the rapidity distribution of top
quarks is broader than that of anti-top quarks due to the difference
in momentum fraction carried by initial-state quarks versus
anti-quarks.  Since the charge asymmetry arises from a
next-to-leading-order effect in quark-antiquark annihilation, while
$t\overline{t}$ pairs are produced at the LHC mainly by gluon fusion,
the predicted asymmetry is even smaller at the LHC than at the
Tevatron. 
  First measurements from ATLAS \cite{ATLAS_afb} with 0.7
$\rm{fb}^{-1}$ and CMS \cite{CMS_afb} with 1.09 $\rm{fb}^{-1}$ come
out consistent with the SM but still with large uncertainties.

\section{Results from the LHC}

As of the end of July, around the time of this conference, ATLAS and
CMS have each collected about 1.5 $\rm{fb}^{-1}$ of data at 7 TeV.
The LHC is running very well, reaching a peak luminosity of about
$2\times 10^{33}$ as of the end of July 2011.
Both ATLAS and CMS have produced search results
with about 1 $\rm{fb}^{-1}$ analyzed.

\subsection{Searches for resonances}

\begin{figure}
\begin{minipage}{0.35\linewidth}
\begin{center}
\includegraphics[width=0.95\linewidth]{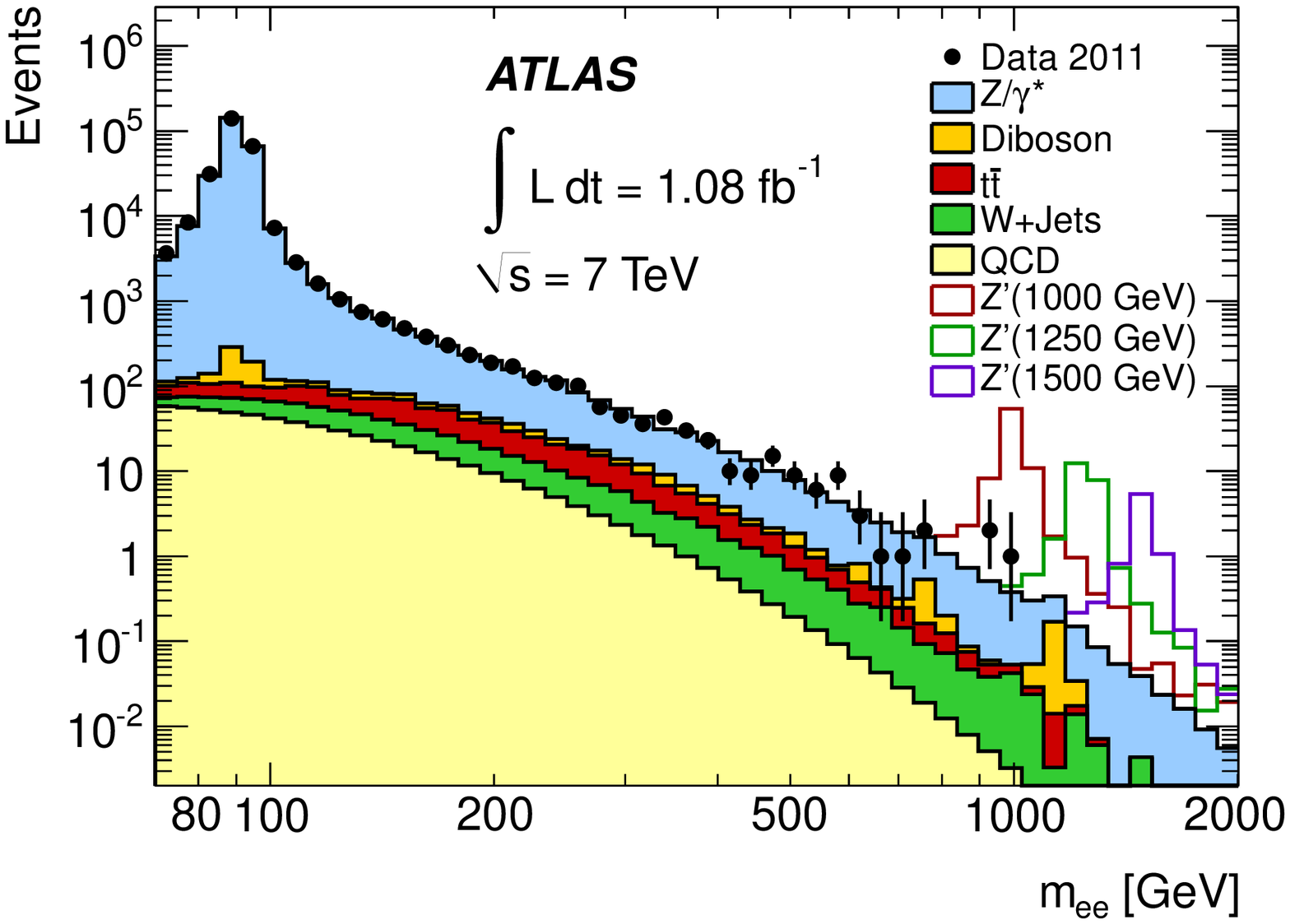}
\end{center}
\end{minipage}
\begin{minipage}{0.275\linewidth}
\begin{center}
\includegraphics[width=0.95\linewidth]{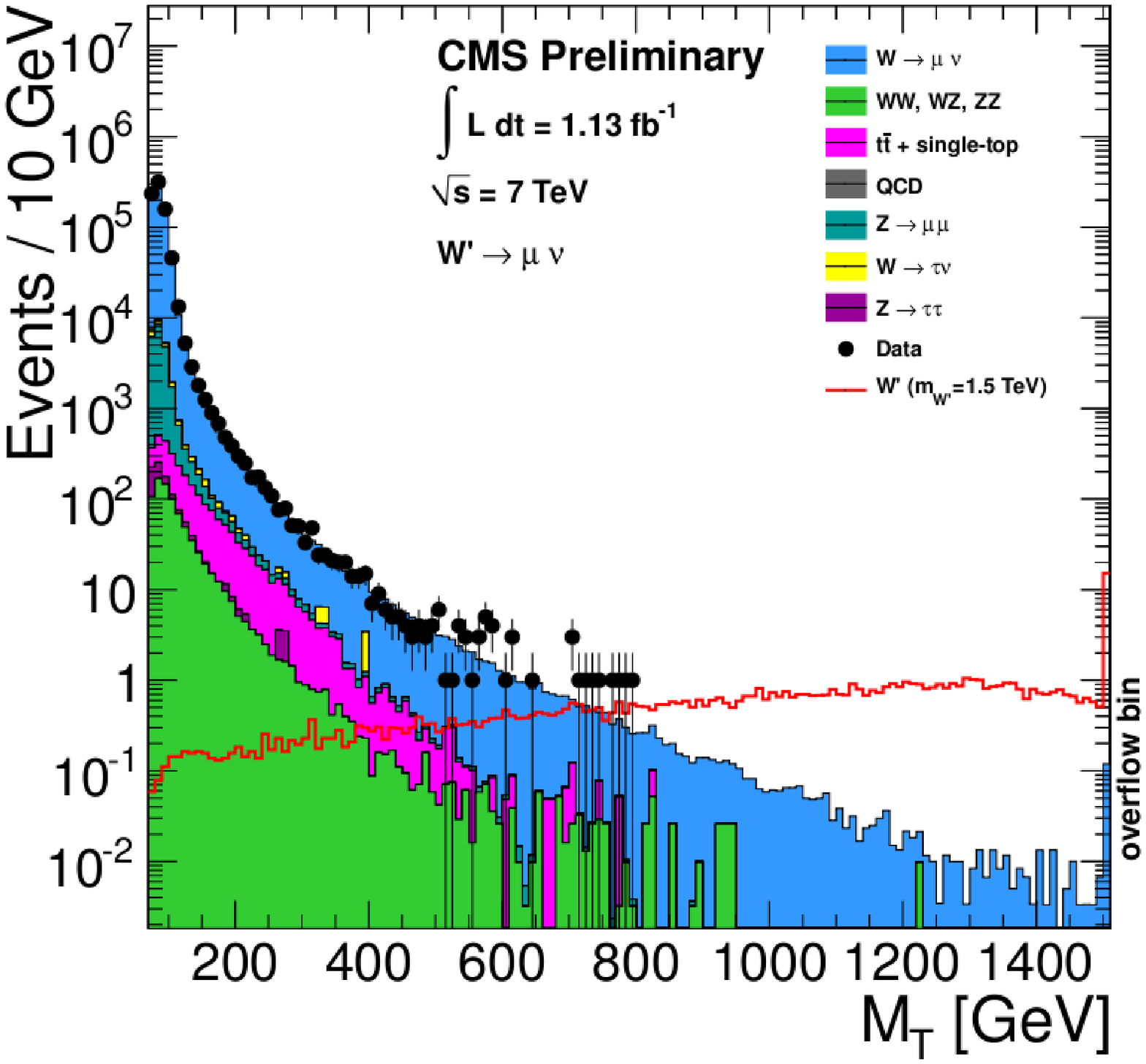}
\end{center}
\end{minipage}
\begin{minipage}{0.35\linewidth}
\begin{center}
\includegraphics[width=0.95\linewidth]{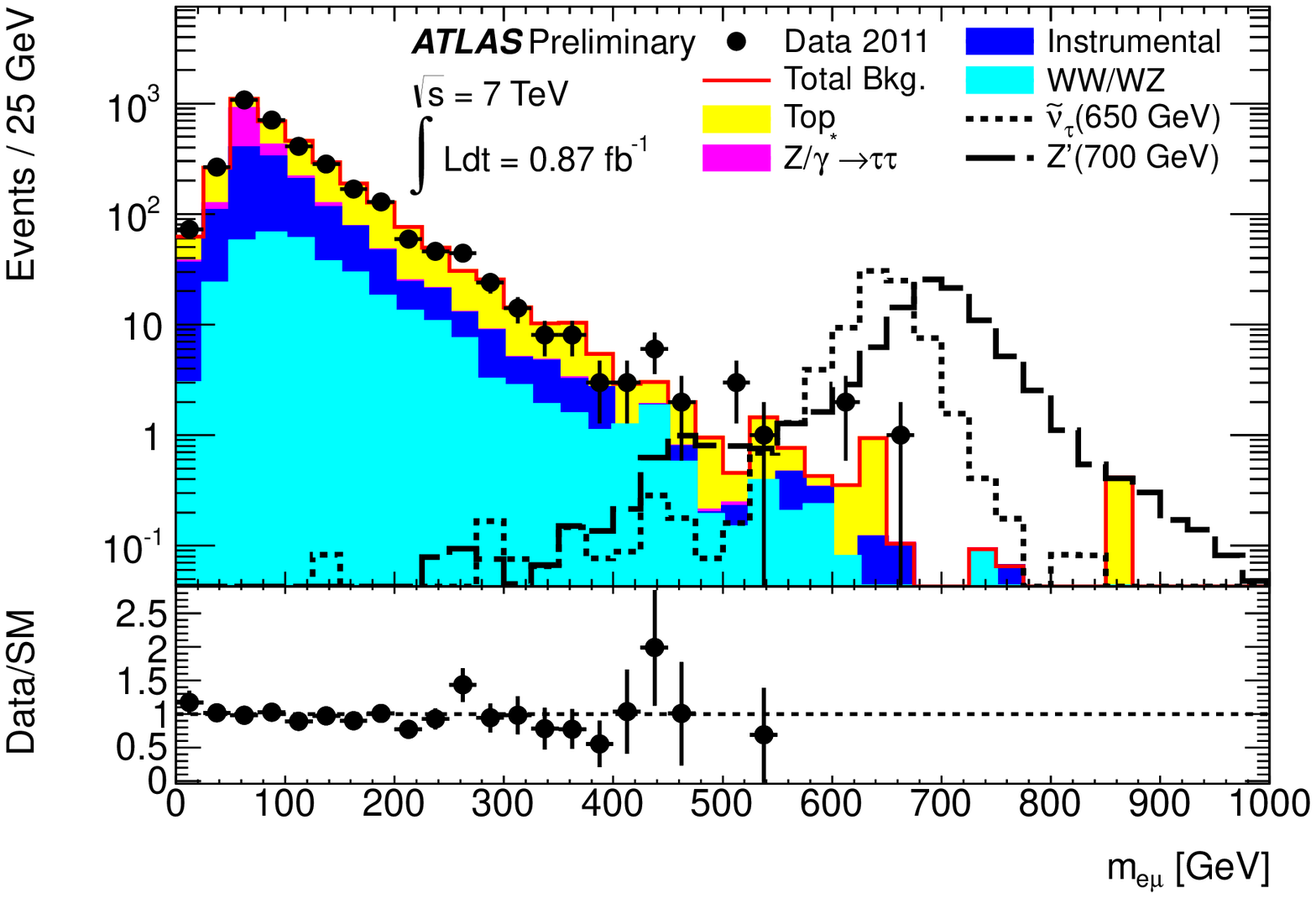}
\end{center}
\end{minipage}
\caption{Left: Dielectron invariant mass
  from ATLAS \cite{ATLAS_Z'}. Center: transverse mass between
  $E_{T}^{miss}$ and the muon from CMS \cite{CMS_W'}. Right:
  $e^\pm\mu^\mp$ invariant mass from ATLAS \cite{ATLAS_emu}.}
\label{fig:resonances}
\end{figure}

We start with the most classic of new physics searches, the dilepton
resonance, looking for bumps in the mass spectrum of isolated high
$p_{T}$
opposite-sign, same-flavor leptons \cite{ATLAS_Z', CMS_Z'}.
ATLAS and CMS have updated their previous  searches to approximately
1.1 $\rm{fb}^{-1}$.
The lepton $p_{T}$ and $\eta$ cuts are
roughly the same in the two experiments.
The background is dominated by the
Z-pole and continuum Drell-Yan production, and is estimated from Monte
Carlo.  As an example, the
dielectron invariant mass
distribution from ATLAS is shown in Fig. \ref{fig:resonances} (left).
No significant deviation
from expectations is seen in either experiment,
and limits are set on a number
of $Z'$ models.  One standard benchmark is the sequential standard model
(SSM) $Z'$ which
has the same couplings and decays as
the SM $Z$ but is heavier.  Both ATLAS and CMS
set a limit\footnote{All limits in this document will be at 95\%
  confidence level, unless otherwise stated.}
of close to 2 TeV on the SSM $Z'$ mass, from
combining $ee$ and $\mu\mu$ channels and now surpass
the limits from the Tevatron.  Limits on other Z' and Randall-Sundrum graviton
models have also been produced that exceed limits from the Tevatron.

ATLAS and CMS have updated their searches
for peaks in the transverse mass ($M_{T}$) spectrum
of events with a high $p_{T}$ lepton and missing transverse
momentum ($E_{T}^{miss}$) \cite{ATLAS_W', CMS_W'}.
The main background comes from the tail of the SM W. This is estimated
by Monte Carlo in the case of ATLAS.  CMS makes a fit to the low
$M_{T}$ region and extrapolates to higher $M_{T}$.  The transverse mass
distribution from CMS in the muon channel is shown in
Fig. \ref{fig:resonances} (center). ATLAS and CMS set
comparable limits of about 2.2 TeV, combining electron- and
muon-channels.  This represents an improvement 
by about 50\%, or approximately 700 GeV,
in the mass limit, for an approximately 30-fold increase in the
integrated luminosity, going from the previous
result based on 35 $\rm{pb}^{-1}$ to the current 1 $\rm{fb}^{-1}$.

ATLAS has also searched for a resonance in isolated $e\mu$
pairs \cite{ATLAS_emu}.
The invariant mass distribution is shown in Fig. \ref{fig:resonances}
(right) where it can be seen that the main background sources
are $t\overline{t}$ and jet instrumental background where a jet from
either $W/Z$+jets or QCD multijet events fakes a lepton.  Monte Carlo
is used to estimate the $t\overline{t}$ background.  The jet
background is estimated using the standard matrix method. In order to
take into account the dependence of the fake rate on the event
kinematics, the matrix equation is solved event by event for a set of
weights; these weights are then summed over all events.
Limits are set on the cross section ($\sigma$) times branching ratio
($BR$), using a R-parity violating model
with a tau sneutrino ($\tilde\nu_\tau$) decaying to $e + \mu$ to
compute the acceptance ($\mathcal{A}$).
Using the same model, limits are also set in the plane
of the $\tilde\nu_\tau$ to $d\overline{d}$ coupling $\lambda'_{311}$
as a function of the $\tilde\nu_\tau$
mass, for various values of the coupling $\lambda_{312}$
of $\tilde\nu_\tau$ to $e\mu$.  ATLAS now
supersedes the D0 limit \cite{D0_emu} on the coupling for almost all masses,
and improves significantly on the limit at high mass.

\begin{figure}
\begin{minipage}{0.32\linewidth}
\begin{center}
\includegraphics[width=0.95\linewidth]{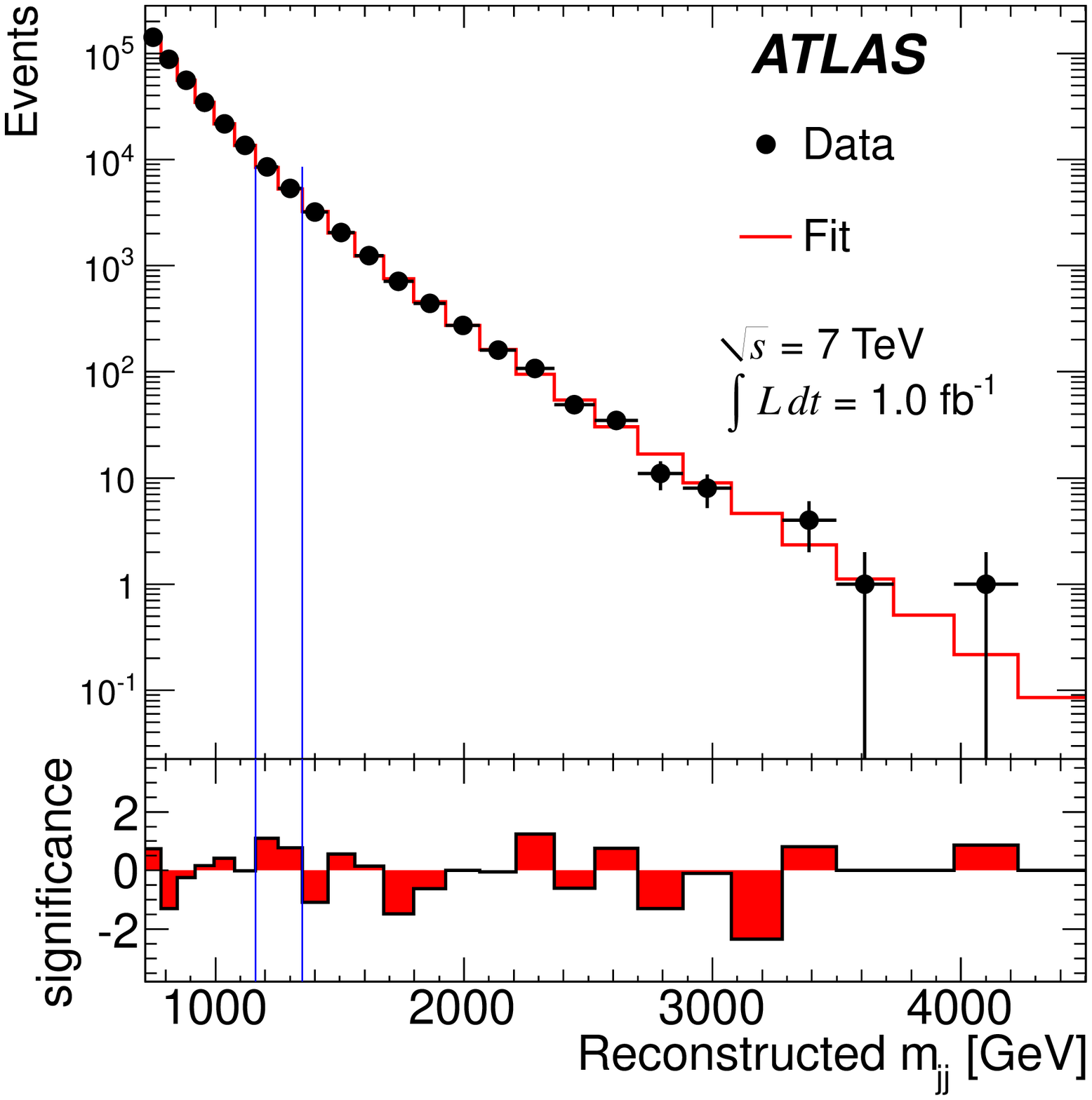}
\end{center}
\end{minipage}
\begin{minipage}{0.32\linewidth}
\begin{center}
\includegraphics[width=0.95\linewidth]{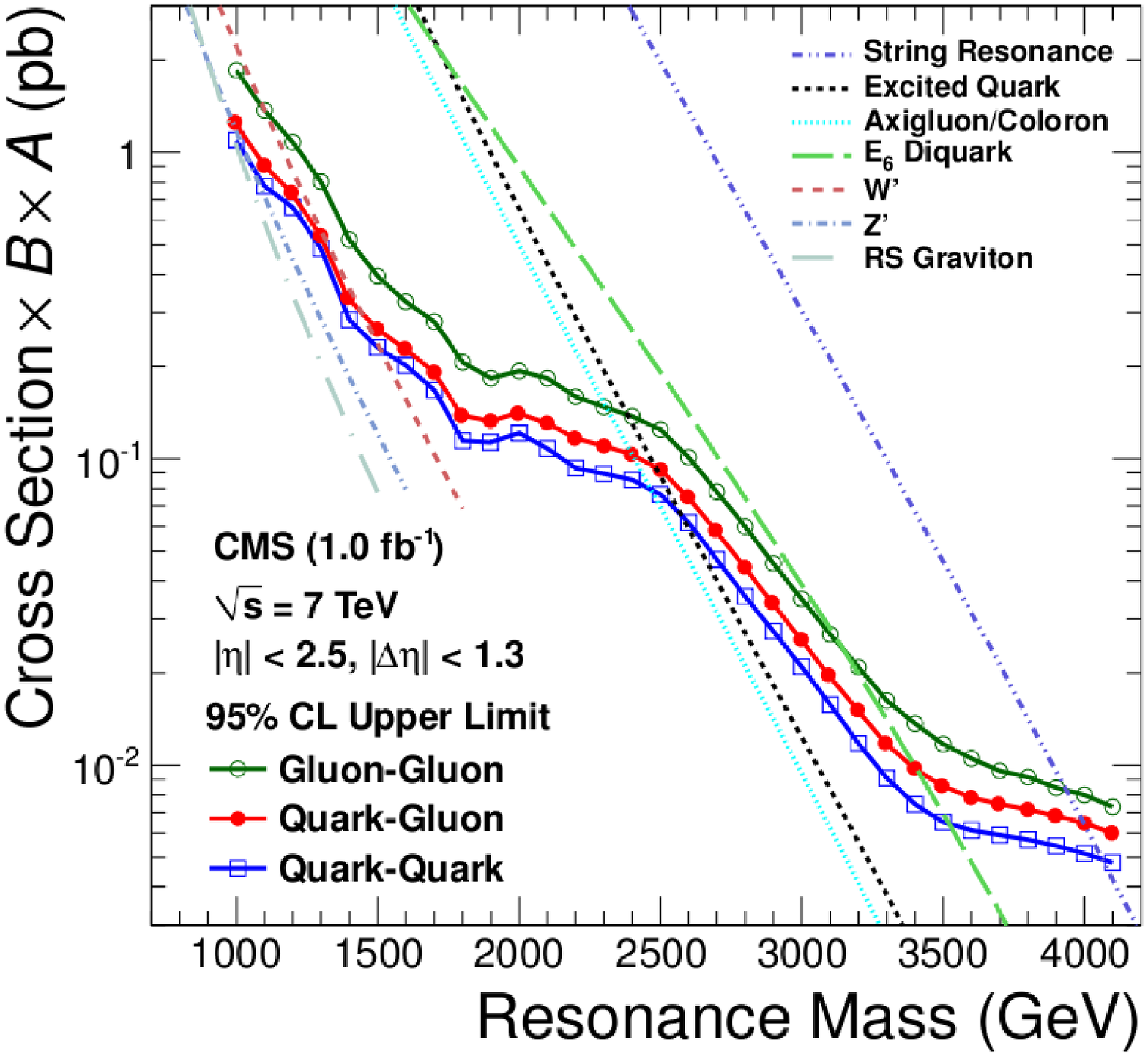}
\end{center}
\end{minipage}
\begin{minipage}{0.32\linewidth}
\begin{center}
\includegraphics[width=0.95\linewidth]{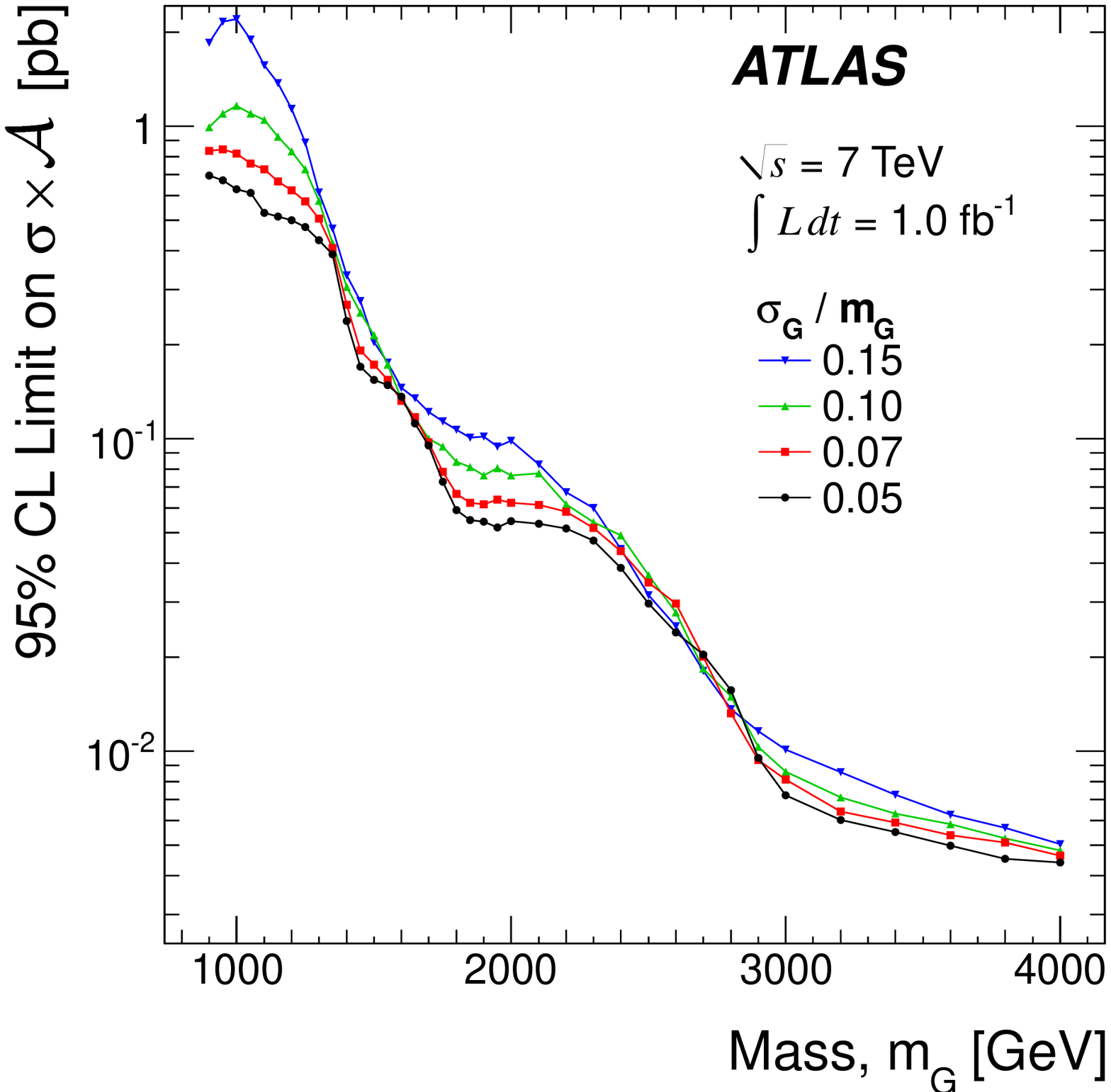}
\end{center}
\end{minipage}
\caption{Left: Dijet invariant mass spectrum from ATLAS
  \cite{ATLAS_mjj}.
  Center/right: Model-independent upper limits from CMS and ATLAS
  on $\sigma \times \mathcal{A}$ for dijet resonances.  Center: 
  For narrow-resonances, the dependence on the parton-level
  subprocess \cite{CMS_mjj}.  Right: The dependence on the resonance
  width \cite{ATLAS_mjj}.}
\label{fig:mjj}
\end{figure}

Another classic search is the hunt for a bump in the dijet mass
spectrum; an example from ATLAS is shown in Fig. \ref{fig:mjj}
(left). ATLAS uses the anti-$k_{t}$ \cite{antikt}
jet algorithm with the distance
parameter $R=0.6$.  CMS starts with anti-$k_{t}$ jets with $R=0.5$,
selects the resulting two highest $p_{T}$ jets and then adds other
jets that were found within a radius of 1.1.
Both ATLAS and CMS derive limits on $\sigma \times
\mathcal{A}$ as a function of the resonance mass \cite{ATLAS_mjj, CMS_mjj}.
Limits in a number of models have been derived.  For example, both
ATLAS and CMS set limits ranging from 2.7 to 3 TeV
on excited quarks and on axigluon models.  The ATLAS mass limits
improve by approximately 30\% in going from the previous results based
on 35 $\rm{pb}^{-1}$ to the current results based on 0.81
$\rm{fb}^{-1}$.  Fig. \ref{fig:mjj} (center) shows model-independent limits
from CMS on
$\sigma \times \mathcal{A}$ for different subprocesses at parton
level, assuming a narrow resonance; the plot on the right from ATLAS
shows the dependence on the width of the resonance.

\begin{figure}
\begin{minipage}{0.45\linewidth}
\begin{center}
\includegraphics[width=0.95\linewidth]{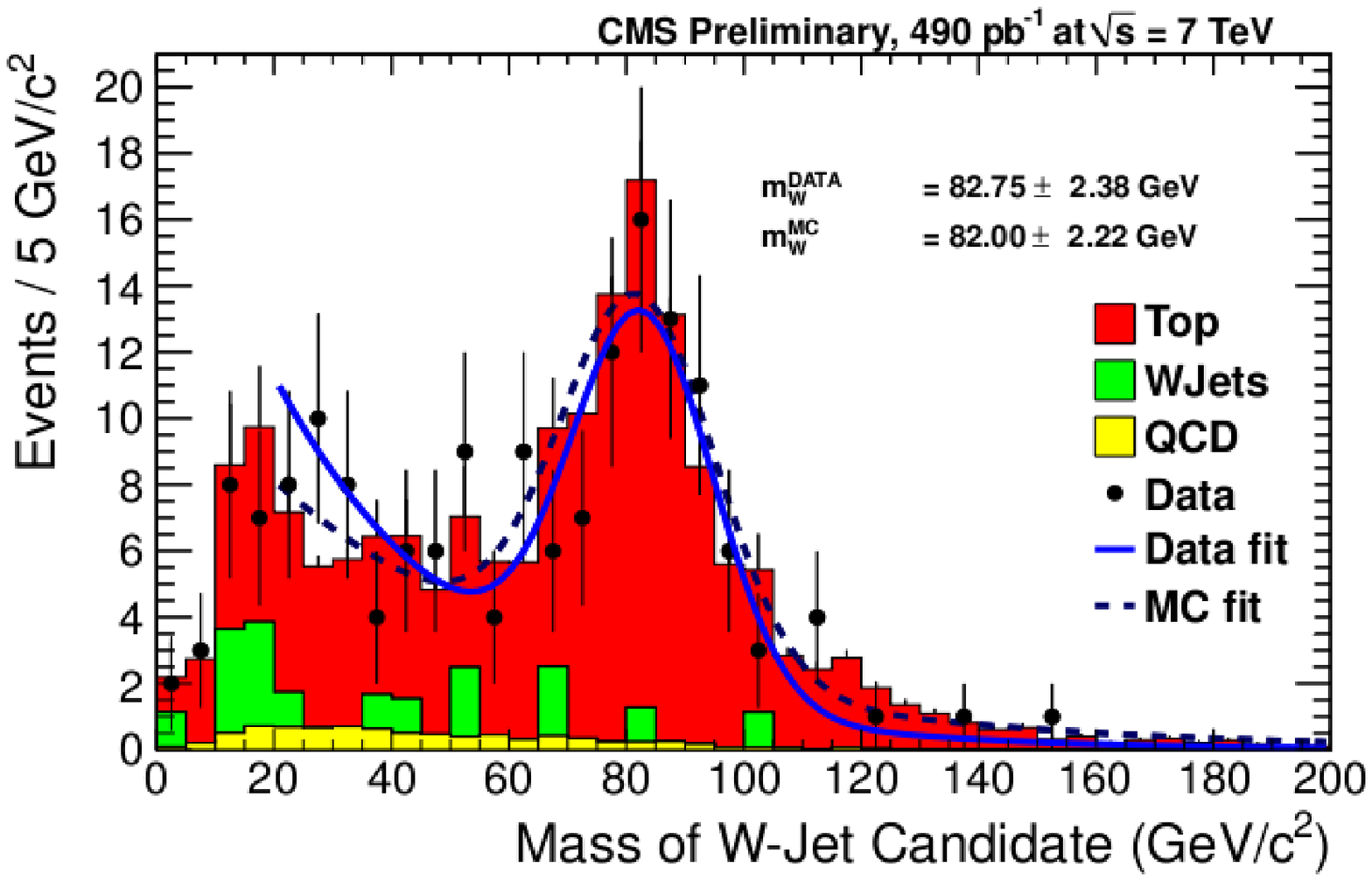}
\end{center}
\end{minipage}
\begin{minipage}{0.45\linewidth}
\begin{center}
\includegraphics[width=0.95\linewidth]{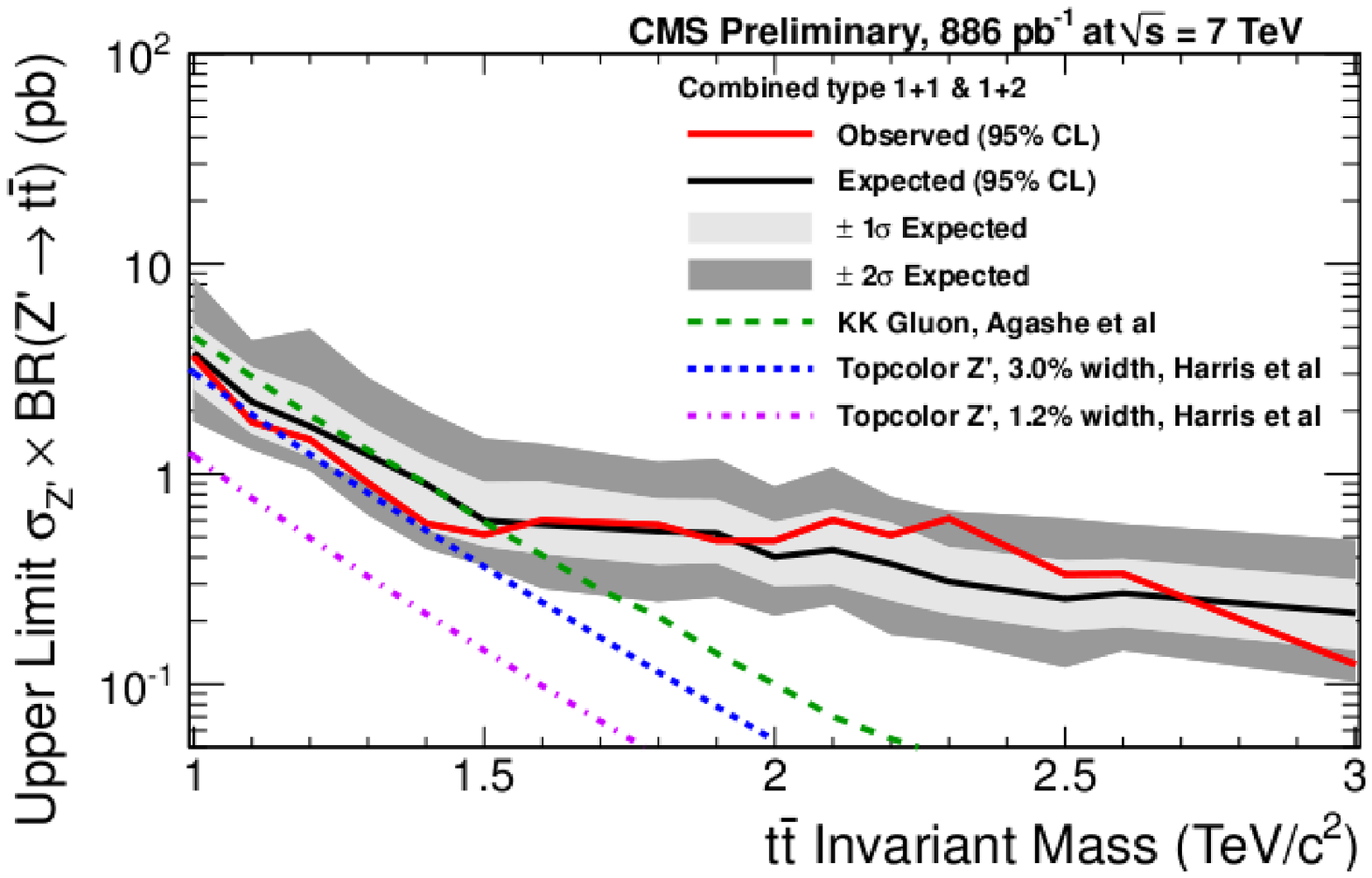}
\end{center}
\end{minipage}
\caption{Left: Mass of the highest mass jet in the hadronic hemisphere
  of boosted semi=leptonic $t\overline{t}$ events.  Right: Limits on
  $\sigma \times BR(Z' \rightarrow t\overline{t})$ from \cite{CMS_tt}.}
\label{fig:tt}
\end{figure}

CMS has updated the search for $t\overline{t}$ resonances
\cite{CMS_tt}, in the all-hadronic channel,
focusing on the case where the top quarks are
sufficiently boosted such that some of the decay products may be
merged; this signature is efficient only for $t\overline{t}$ masses
above approximately 1 TeV.  Two topologies are considered: i) a dijet
topology in which all three jets from the top quark are merged, and
ii) a 3-jet topology in which the decay products are completely merged
in one hemisphere, while only the decay products of the W are merged
in the other.  Top quarks and W's are tagged by analyzing the jet
substructure.  The jet energy scale for merged jets is checked by
examining the mass of the highest
mass jet in the hadronic hemisphere of boosted
semileptonic $t\overline{t}$ events.  Fig. \ref{fig:tt} (left) shows
the position of the resulting W-mass peak.  
The background, dominated by QCD multijets, is
estimated directly from the data, by selecting events with one
top-tagged jet, and applying a top-mistag rate, also estimated from
the data.  The small background from $t\overline{t}$ is estimated via
Monte Carlo.  Limits are set on the production cross section times
branching ratio for a $Z' \rightarrow t\overline{t}$, as shown in
Fig. \ref{fig:tt} (right).

\subsection{Other searches}

CMS has searched for a fourth-generation $t'$ quark \cite{CMS_tprime},
assuming a decay
to $W+b$.  The search is done in the semi-leptonic channel, requiring
one isolated lepton and four or more jets, at least one of which is
b-tagged.  A kinematic fit is performed to compute the $t'$ mass, and
the search is performed in the plane of $H_{T}$ versus the fitted
mass, where $H_{T}$ is defined as the scalar sum of $E_{T}^{miss}$ and
the transverse
energies of the lepton and jets.   The
main background comes from $t\overline{t}$ and to a lesser extent
$W$+jets; all backgrounds are estimated using the Monte Carlo.
Combining electron and muon channels, $t'$ quarks are excluded below
450 GeV, surpassing the Tevatron limit.

\begin{figure}
\begin{minipage}{0.495\linewidth}
\begin{center}
\includegraphics[width=0.95\linewidth]{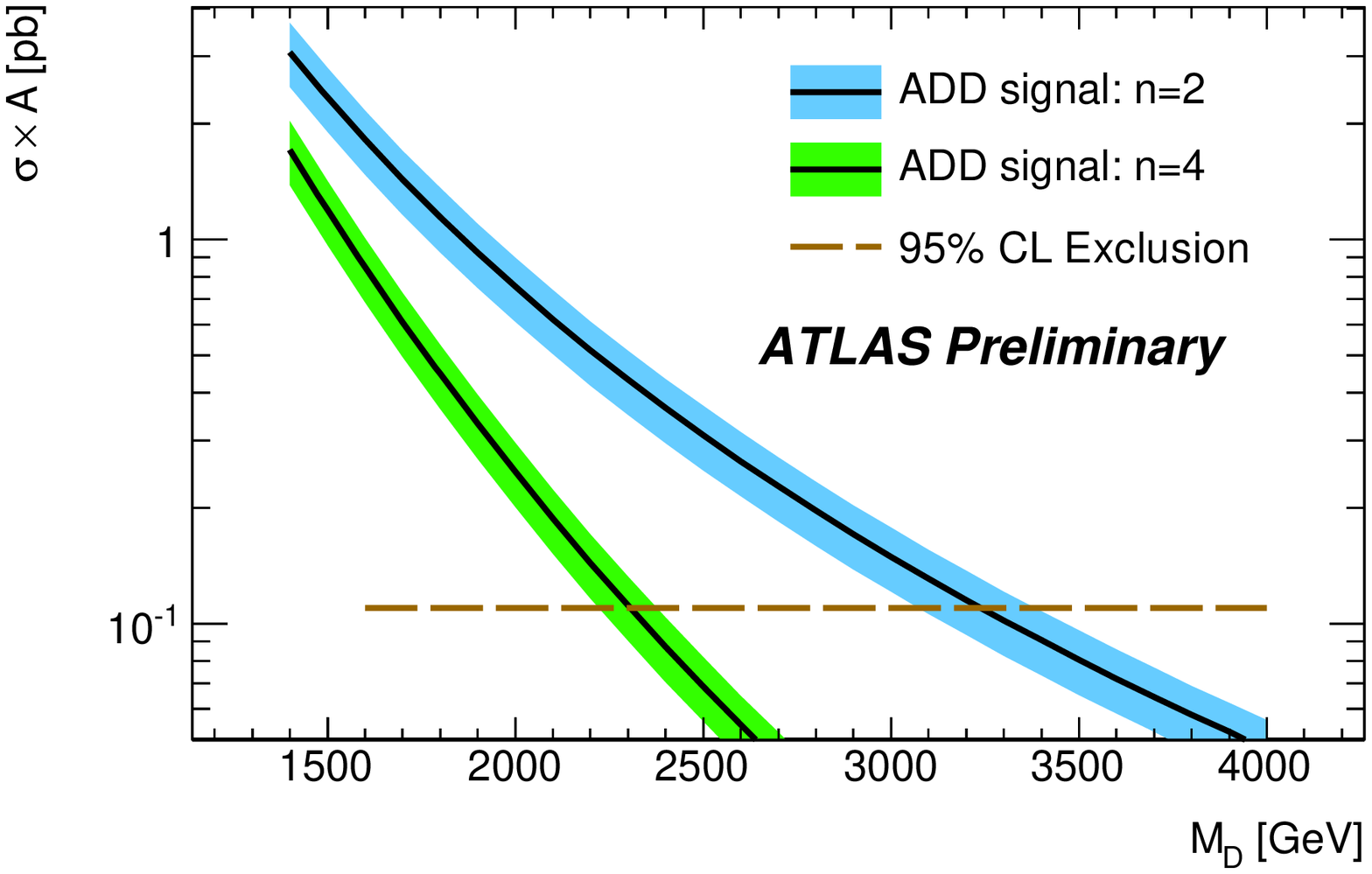}
\end{center}
\end{minipage}
\begin{minipage}{0.495\linewidth}
\begin{center}
\includegraphics[width=0.95\linewidth]{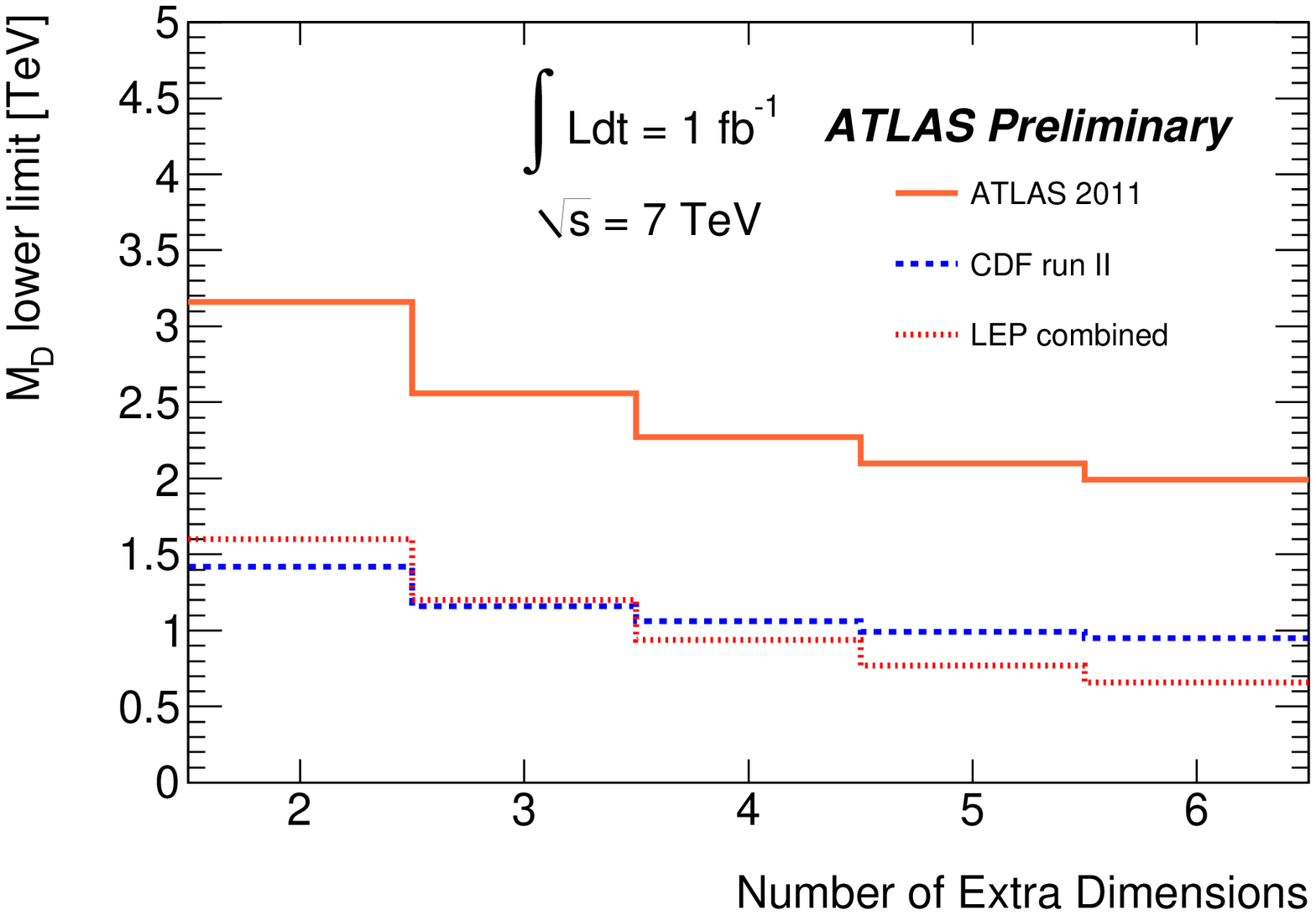}
\end{center}
\end{minipage}
\caption{Left: Model-independent limit on $\sigma \times \mathcal{A}$
  as a function of the $4+n$-dimensional Planck scale, $M_{D}$ for 2
  and 4 extra dimensions.  Right: Limits on $M_{D}$ as a function of
  the number of extra dimensions.  From \cite{ATLAS_monojet}.}
\label{fig:ATLAS_monojet}
\end{figure}

ATLAS has updated the search for monojets and large $E_{T}^{miss}$ \cite{ATLAS_monojet}.
Three signal regions are defined, with varying cuts on the leading jet
$p_{T}$, $E_{T}^{miss}$ and the veto thresholds for additional jets in
the event.  The dominant backgrounds from $Z(\rightarrow
\nu\overline\nu)$+jets and from $W(\rightarrow \ell\nu)$+jets are
estimated using Monte Carlo normalized to
a muon control sample (but with the requirements on
jet $p_{T}$, $E_{T}^{miss}$ and subleading jet vetoes the same as in
the signal search)  from the data.  Multijet background is estimated
by a linear extrapolation of the $p_{T}$ of the subleading jet below the
jet veto threshold, using a sample of events where $E_{T}^{miss}$
points along the subleading jet direction.
Fig. \ref{fig:ATLAS_monojet} (left) shows the resulting model-independent limits
on $\sigma \times \mathcal{A}$.
Limits are also set in the ADD
large extra dimensions model on the $4+n$-dimensional Planck scale
($M_{D}$) as a function of the number of extra dimensions.  For the
number of extra dimensions ranging from 2 to 6, $M_{D}$ values between
3.2 TeV and 2.0 TeV are excluded as shown in Fig. \ref{fig:ATLAS_monojet}
(right).

\begin{figure}
\begin{minipage}{0.32\linewidth}
\begin{center}
\includegraphics[width=0.95\linewidth]{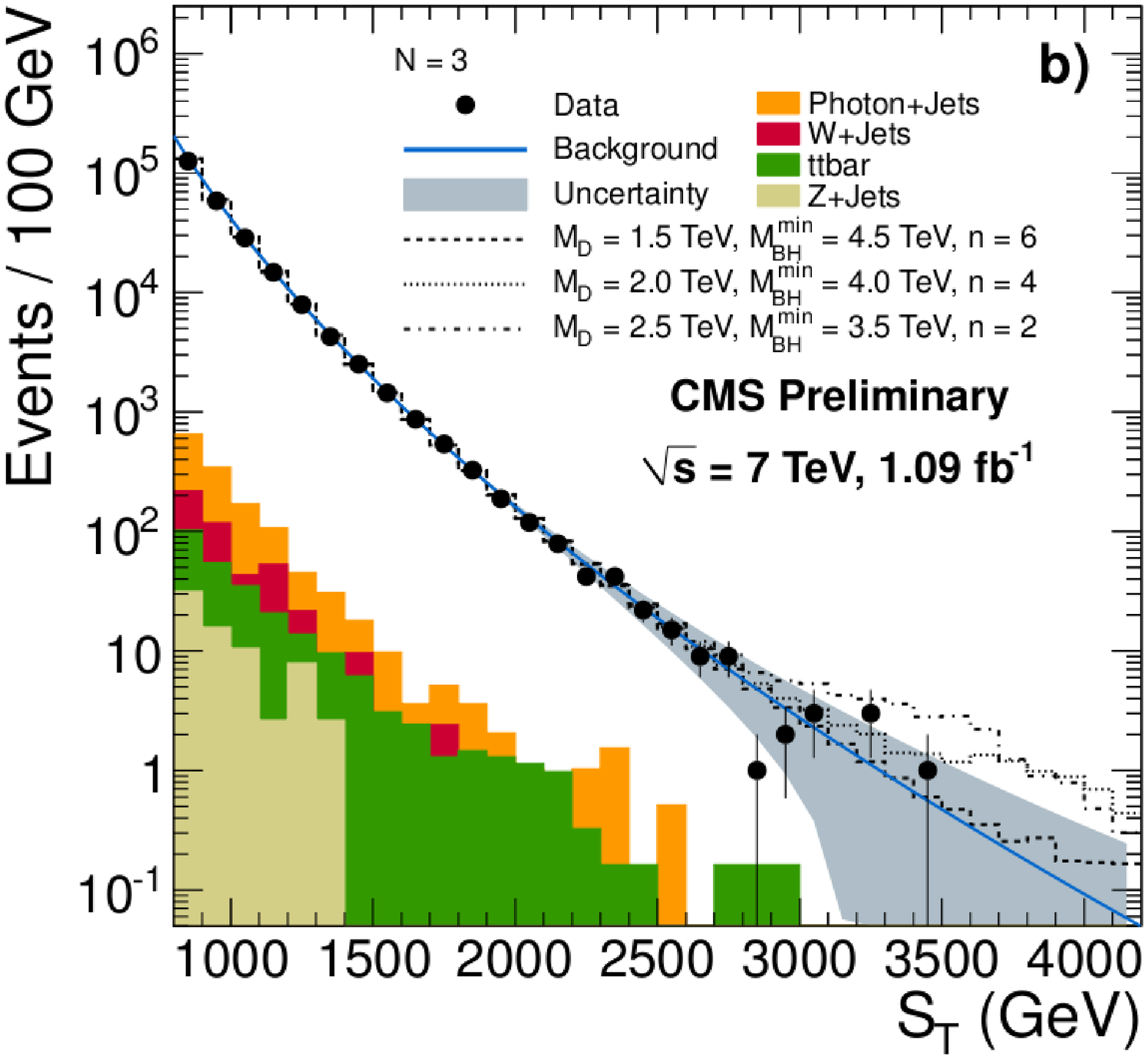}
\end{center}
\end{minipage}
\begin{minipage}{0.32\linewidth}
\begin{center}
\includegraphics[width=0.95\linewidth]{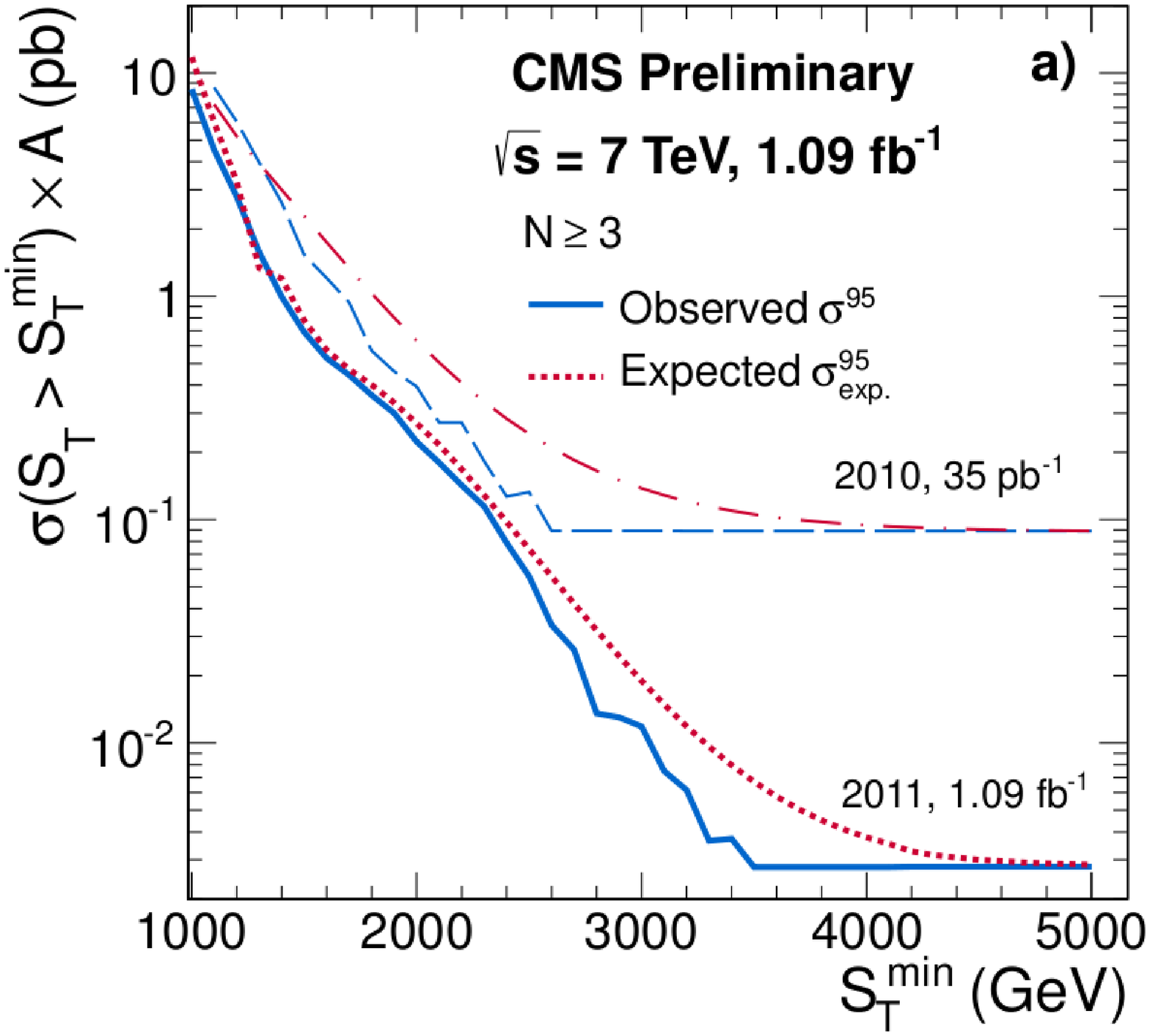}
\end{center}
\end{minipage}
\begin{minipage}{0.32\linewidth}
\begin{center}
\includegraphics[width=0.95\linewidth]{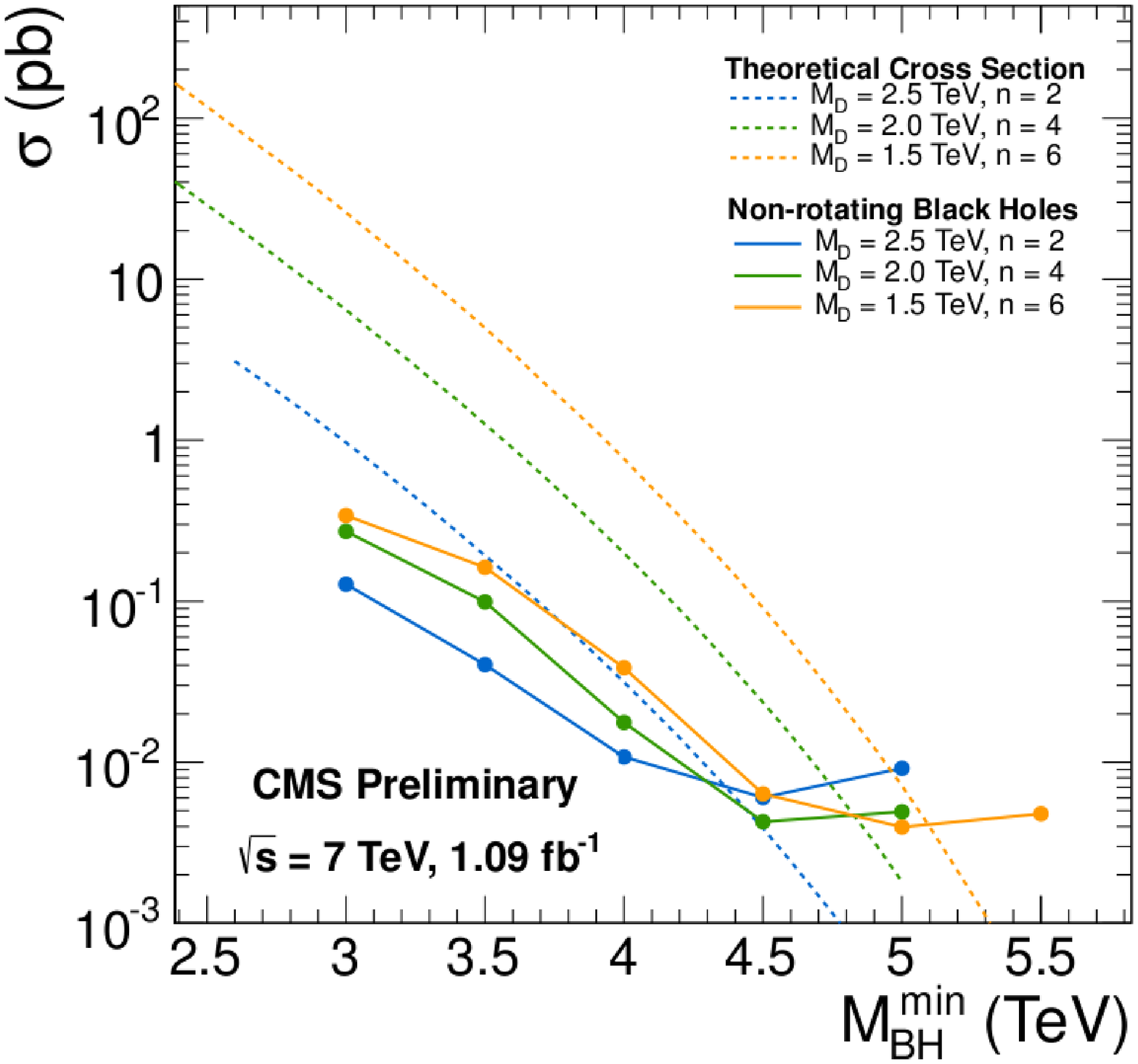}
\end{center}
\end{minipage}
\caption{Left: Total transverse energy for events with $N=3$ photons,
  electrons, muons or jets in the final state.
  Center: Model-independent limits on $\sigma \times \mathcal{A}$
  for $S_{T} > S_{T}^{min}$ as a function of $S_{T}^{min}$,  for
  events with 3 or more objects in the event.  Righ: Cross section
  limits compared with signal production cross sections from the
  BLACKMAX generator.  The colored solid lines show the experimental
  cross section limits, while the dotted lines sohw the corresponding
  signal cross sections.  From \cite{CMS_BH}.}
\label{fig:BH}
\end{figure}

CMS has updated the search for microscopic black holes \cite{CMS_BH}.
The main discriminating variables are: i)  $S_{T}$, the scalar sum of the
$p_{T}$ of all jets, electrons, muons, photons and $E_{T}^{miss}$
where each of the objects is required to be greater than 50 GeV, and
ii) $N$, the number of objects in the event.  The main background of
multijet and photon+jets events is estimated by exploiting the
independence of the shape of the $S_{T}$ distribution on the
multiplicity of objects in the event; the $S_{T}$ shape is determined
for $N=2$ where no signal is present, as verified in the dijet
analysis.  The distribution of $S_{T}$ for $N=3$ is shown in
Fig. \ref{fig:BH} (left).
Limits are set on $\sigma \times \mathcal{A}$ for $S_{T} >
S_{T}^{min}$, as shown in Fig. \ref{fig:BH} (center) for $N \ge 3$.
Limits are also
set on black hole masses in the context of the BLACKMAX
\cite{Blackmax} model, as
shown in Fig. \ref{fig:BH} (right); black hole masses below
approximately  4.5-5
TeV are excluded in this model.

\subsection{Searches for slow-moving or stopped charged, massive particles}

\begin{figure}
\begin{minipage}{0.31\linewidth}
\begin{center}
\includegraphics[width=0.95\linewidth]{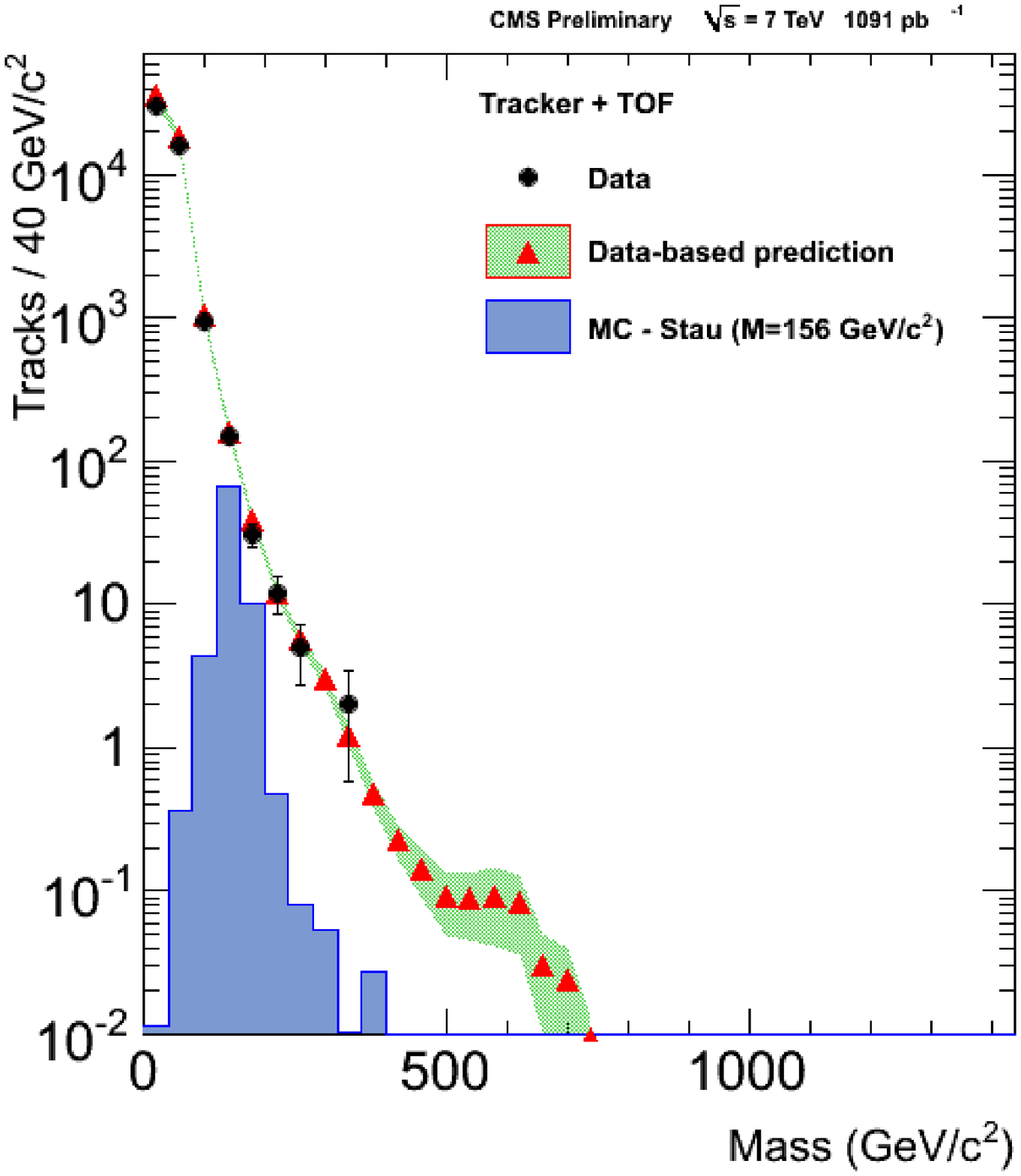}
\end{center}
\end{minipage}
\begin{minipage}{0.33\linewidth}
\begin{center}
\includegraphics[width=0.95\linewidth]{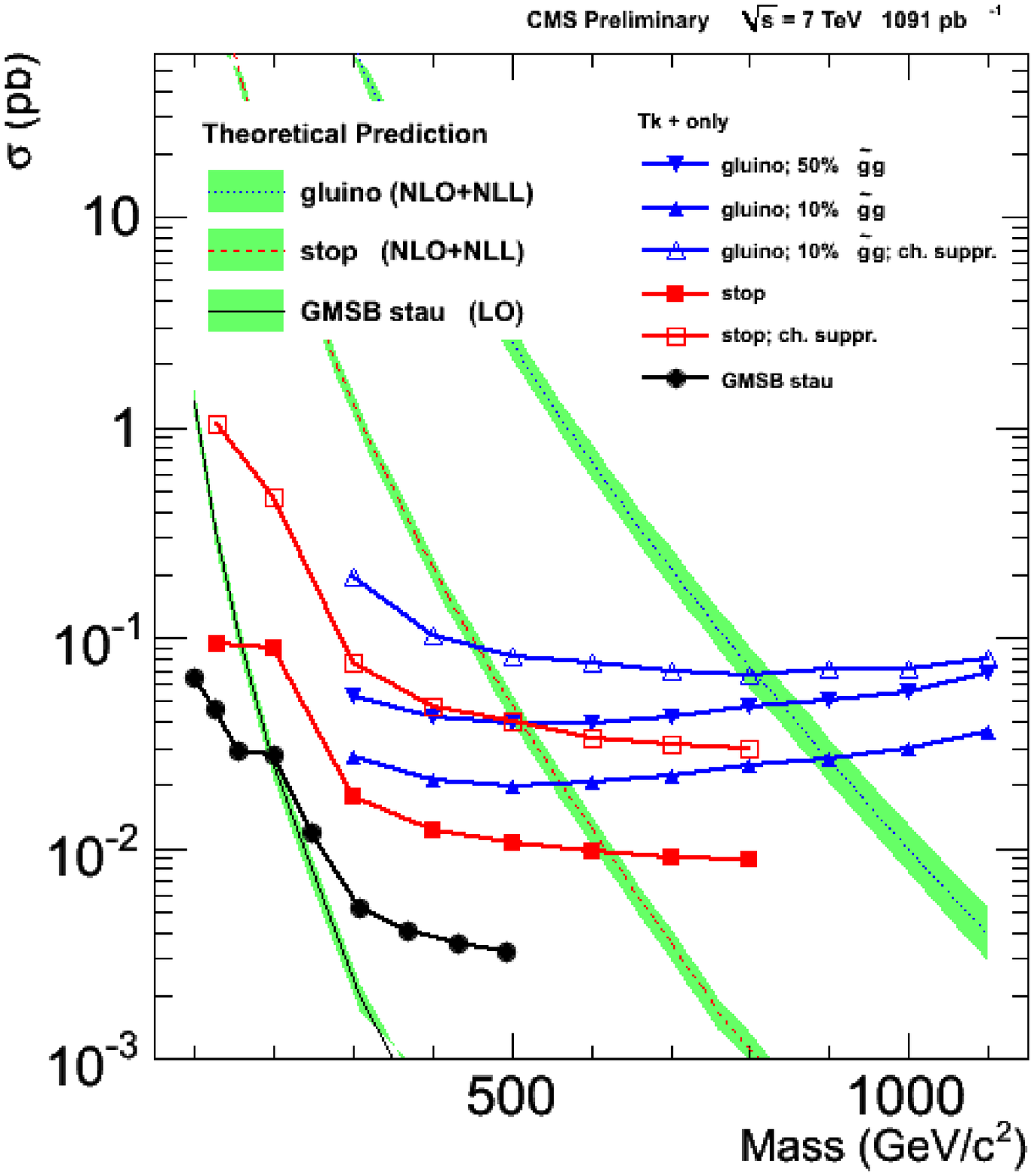}
\end{center}
\end{minipage}
\begin{minipage}{0.33\linewidth}
\begin{center}
\includegraphics[width=0.95\linewidth]{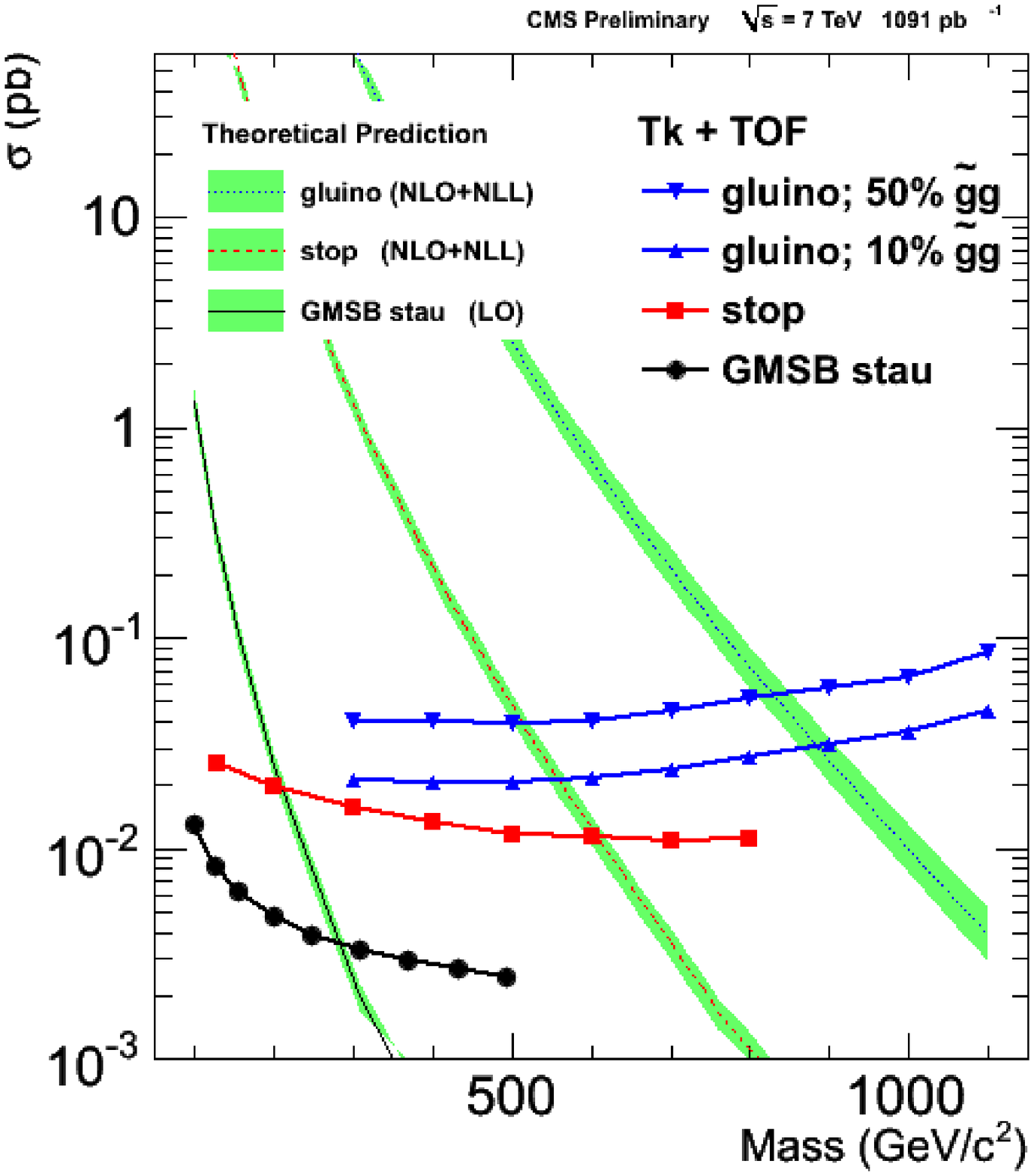}
\end{center}
\end{minipage}
\caption{Left: Mass distribution from the combined analysis of
  tracker dE/dx and muon
  TOF \cite{CMS_LLP}. Center/right: Theoretical cross sections and
  observed limits on the cross section for a variety of models.
  Center: tracker dE/dx analysis.  Right: combined tracker dE/dx
  plus muon TOF
  analysis.}
\label{fig:LLP}
\end{figure}

CMS has updated the search for slow-moving, massive, charged
particles \cite{CMS_LLP}.
These arise quite generically in many SUSY models, for
example.  They make two types of measurements, tracker-only
and tracker plus the
muon system.  An estimation of the mass comes from the track momentum
and either the track dE/dx or the time-of-flight (TOF) to the muon system.
The background is estimated via the commonly called ``ABCD'' method,
exploting the lack of correlation between momentum, TOF and dE/dx.
The resulting mass distribution from the combined analysis is shown in
Fig. \ref{fig:LLP} (left).
From the tracker only analysis, they set limits on stop R-hadrons and
gluino R-hadrons, reaching limits close to 1 TeV for the gluino case,
as shown in Fig. \ref{fig:LLP} (center).
From the combined analysis, shown in Fig. \ref{fig:LLP} (right)
they exclude the partners of the tau lepton (staus) in a minimal GMSB model
(following SPS line 7 \cite{SPS7}), excluding masses below 293 GeV.

CMS has also updated the search for stopped gluino/stop R-hadrons
\cite{CMS_stopped}.
They
look for the R-hadron decay in the calorimeter in between bunches or
during interfill periods.
Potential background sources include detector noise, beam-related background,
and cosmics.  These are estimated using 2010 data.  Noise and cosmics
rates were measured when
the LHC luminosity was around $10^{28}$ $\rm{cm}^{-2}\rm{sec}^{-1}$.  The
rate from beam background and noise+cosmics was estimated from the rest of
the 2010 run. Given vastly different
beam conditions in 2011 and potentially different detector conditions,
this background estimation might be subject to some uncertainty,
but in the end the CMS observations match expectations over a wide
range of R-hadron lifetimes.  They set mass limits on stop and gluino
R-hadrons of 337 and 601 GeV, respectively, for
lifetimes between 10 ps and 1000 seconds, and assuming direct decay to
the lightest supersymmetric particle (LSP).

\subsection{Searches for SUSY}

Both ATLAS and CMS
have updated their search for SUSY in the jets + $E_{T}^{miss}$
channel \cite{ATLAS_0l, CMS_0l}.  ATLAS explores 5
signal regions, based on differing jet multiplicities and different
cuts on $E_{T}^{miss}$ and $H_{T}$, the scalar sum of the $E_{T}$ of
the jets in the event.  CMS does an analysis in the $\alpha_{T}$
variable, defined for a 2 jet topology as $\alpha_{T} =
E_{T}^{jet2}/M_{T}$ where $M_{T}$ is the transverse mass of the two
jets; events with higher jet
multiplicities are treated by merging the jets into two mega jets.
Also of note is that ATLAS does a cut and count analysis, while CMS
exploits for the first time shape information.

Backgrounds
are estimated in both experiments with a mix of data-driven and Monte
Carlo based methods.  QCD multijet background is expected to be
negligible in both experiments; background reduction is achieved
through cuts on the azimuthal angle between jets and $E_{T}^{miss}$,
on $E_{T}^{miss}$/$H_{T}$, and in the case of CMS, on $\alpha_{T}$.
The small QCD multijet background is confirmed in ATLAS by a
dedicated study in which well-measured multijet events are smeared by
jet energy response functions that have been separately measured in
the data in three-jet events where two well-measured jets recoil
against a third jet which points along the $E_{T}^{miss}$ direction.
CMS estimates $W$+jets background together with
$t\overline{t}$ by a muon control sample, primarily selected by a
requirement on $M_{T}$; the background in the signal region is
estimated by scaling the number of events in the control sample by a
factor determined from Monte Carlo.  In the ATLAS analysis, $W$+jets
and $t\overline{t}$ backgrounds are estimated separately, based again
on a lepton control sample selected with $E_{T}^{miss}$ and $M_{T}$
cuts, but using a b-tag requirement (veto) to select the
$t\overline{t}$ ($W$+jets) sample; as in the CMS analysis,
extrapolation to the signal region
is performed via scaling factors derived from Monte Carlo.
$Z(\rightarrow \nu\overline\nu)$+jets background is determined in both
experiments via a control sample of $\gamma$+jets events; ATLAS also
uses control samples of $Z(\rightarrow \ell^{+}\ell^{-})$+jets.
Background systematics are approximately comparable for the two
experiments. 
ATLAS quotes a 50\%
uncertainty on $t\overline{t}$ background,
but the actual level of $t\overline{t}$ background is fairly
small. CMS quotes a 30\%
uncertainty on $W$+jets and $t\overline{t}$ combined,
while ATLAS quotes about 25\% for $W$+jets.
For $Z$+jets, ATLAS has a roughly 20\% uncertainty,
to be compared to 40\% for CMS.

\begin{figure}
\begin{minipage}{0.45\linewidth}
\begin{center}
\includegraphics[width=0.95\linewidth]{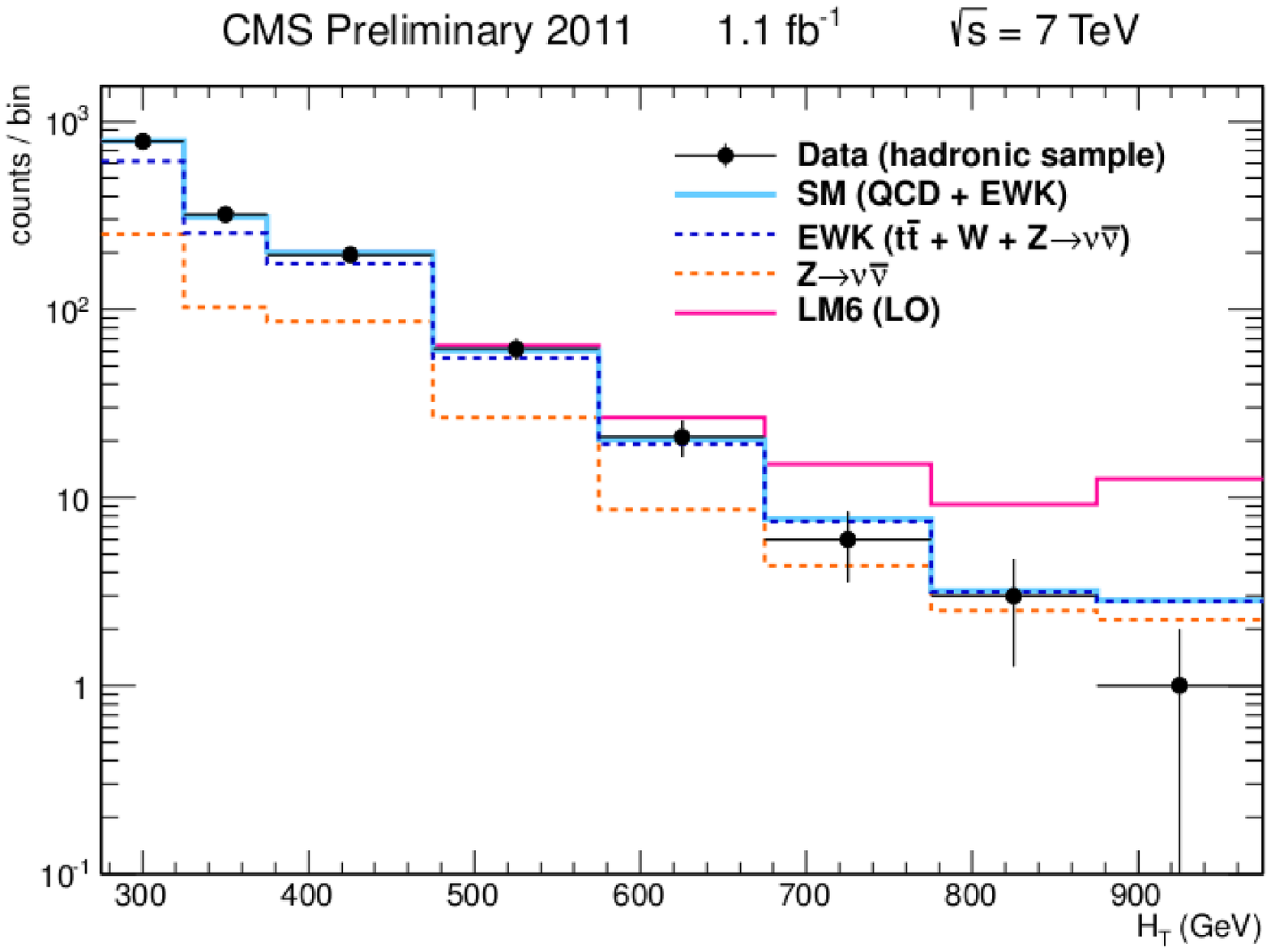}
\end{center}
\end{minipage}
\begin{minipage}{0.45\linewidth}
\begin{center}
\includegraphics[width=0.95\linewidth]{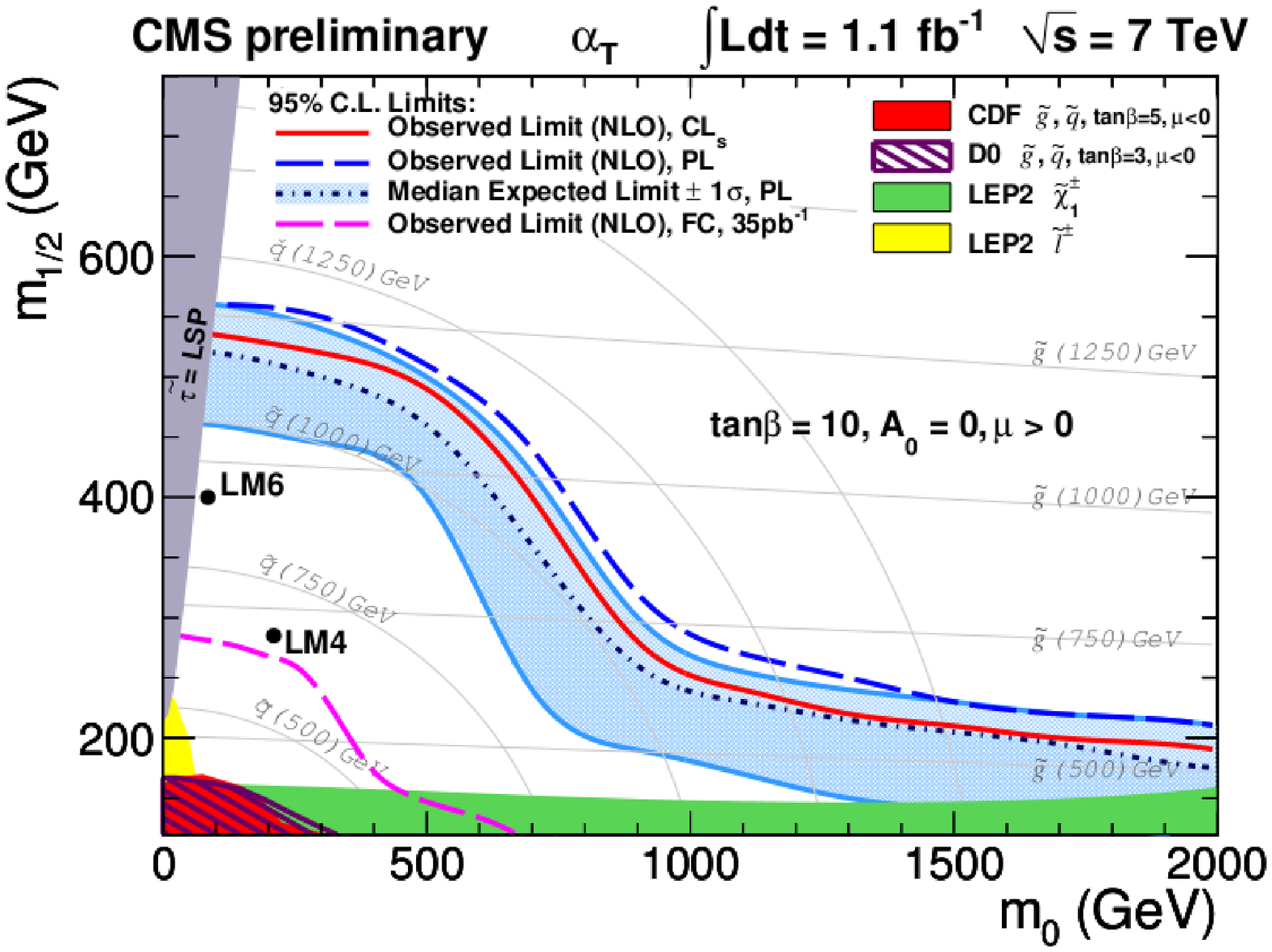}
\end{center}
\end{minipage}
\caption{Left: $H_{T}$ distribution from CMS the SUSY search based on the
  $\alpha_{T}$ variable.  Right: SUSY limits in the MSUGRA/CMSSM
  model. From \cite{CMS_0l}.}
\label{fig:0leptonCMS}
\end{figure}

\begin{figure}
\begin{minipage}{0.45\linewidth}
\begin{center}
\includegraphics[width=0.95\linewidth]{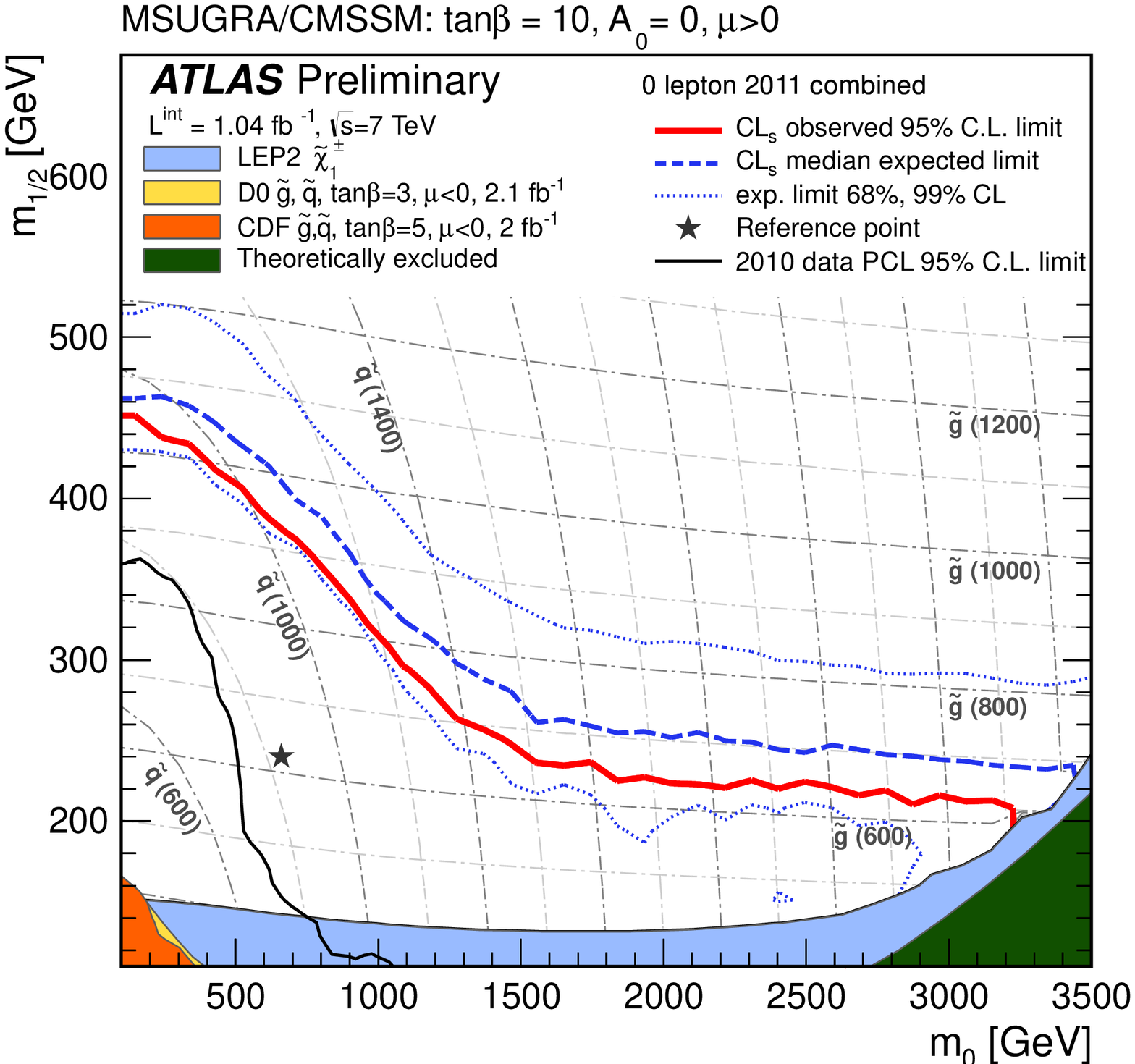}
\end{center}
\end{minipage}
\begin{minipage}{0.45\linewidth}
\begin{center}
\includegraphics[width=0.95\linewidth]{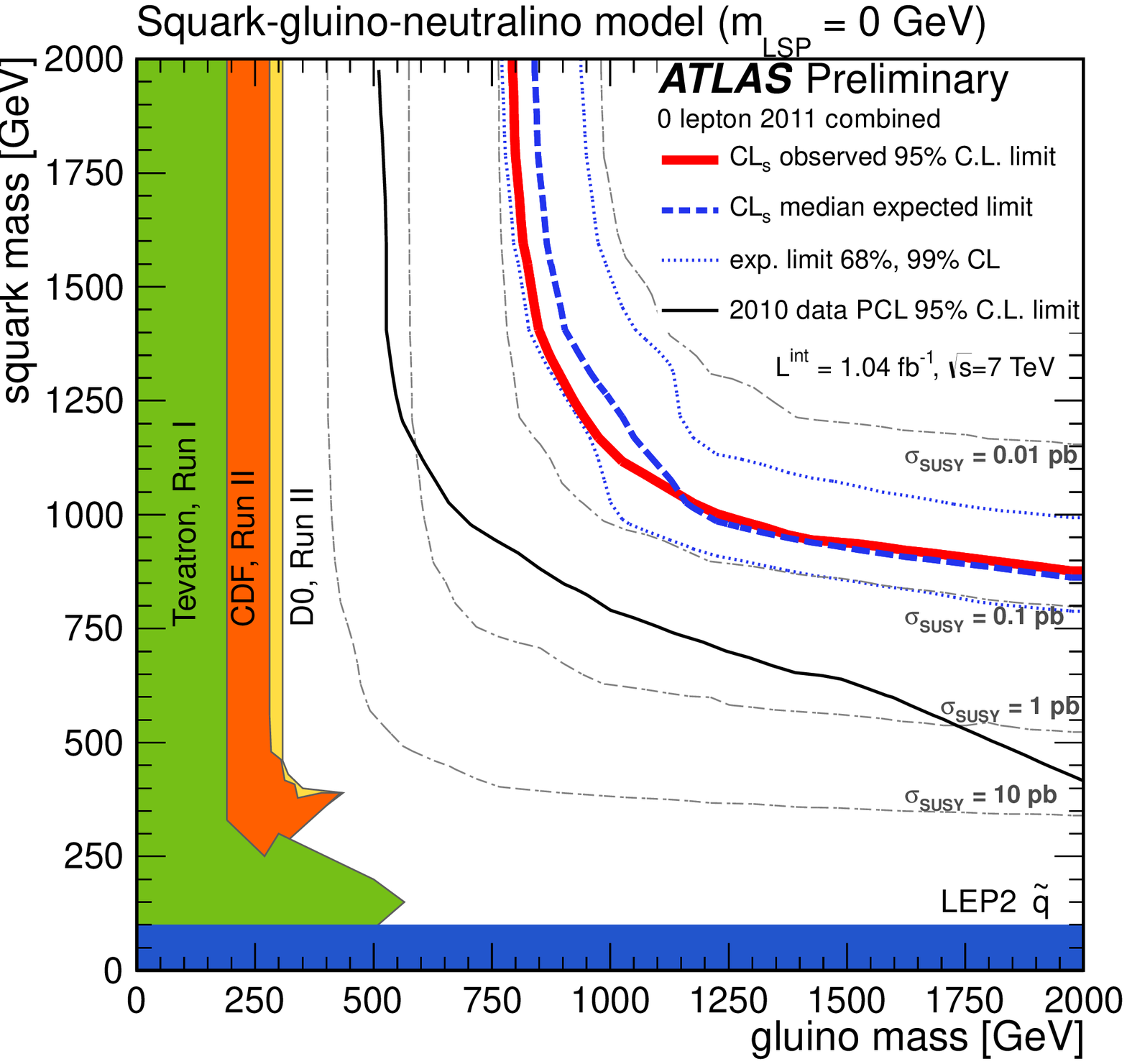}
\end{center}
\end{minipage}
\caption{SUSY limits from ATLAS 
  \cite{ATLAS_0l}.   Left: in the MSUGRA/CMSSM model. Right: gluino and
  squark mass limits in a simplified MSSM model, containing
  only the gluino, first- and second-generation squarks and a massless
  LSP.}
\label{fig:0leptonATLAS}
\end{figure}

In both experiments, observations are consistent with expectations
from the SM.
For example, Fig. \ref{fig:0leptonCMS} (left) shows the $H_{T}$
distribution from CMS compared to expectations. 
Fig. \ref{fig:0leptonCMS} (right)  shows the limits obtained by CMS
in the MSUGRA/CMSSM model.
The corresponding limits from ATLAS are shown in
Fig. \ref{fig:0leptonATLAS} (left).
Both
experiments are setting limits on gluino and squark masses approaching
1 TeV in certain regions of the parameter space.  As shown in
Fig. \ref{fig:0leptonATLAS} (right) ATLAS has also
updated the limits in a simplified MSSM model containing gluinos,
squarks of the first and second generation and the LSP, with all other
sparticles set to very high mass.  Limits approach 800 GeV for gluino
masses, and are insensitive to the value of the LSP mass for LSP
masses up to approximately 200 GeV.

\begin{figure}
\begin{minipage}{0.495\linewidth}
\begin{center}
\includegraphics[width=0.95\linewidth]{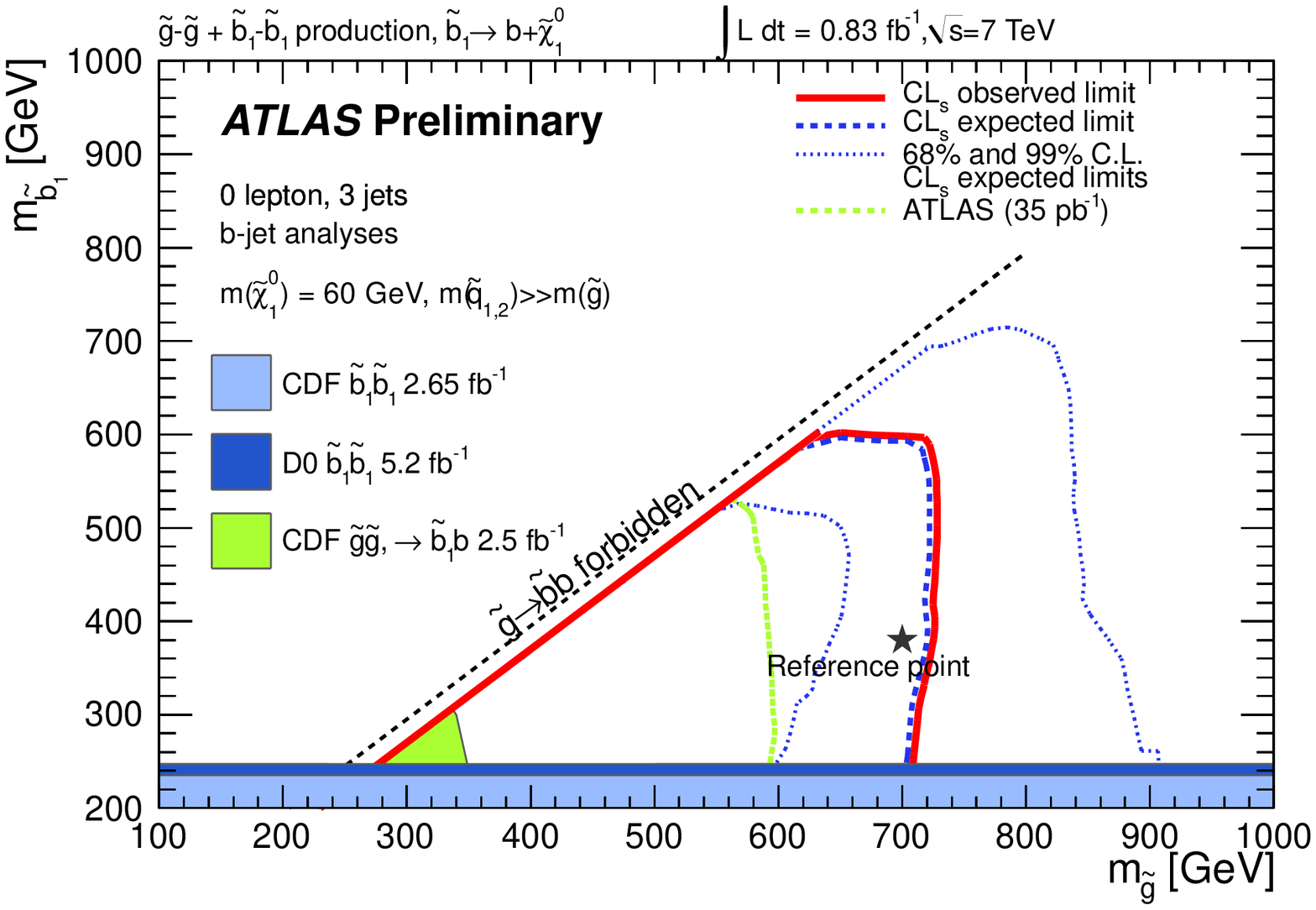}
\end{center}
\end{minipage}
\begin{minipage}{0.495\linewidth}
\begin{center}
\includegraphics[width=0.95\linewidth]{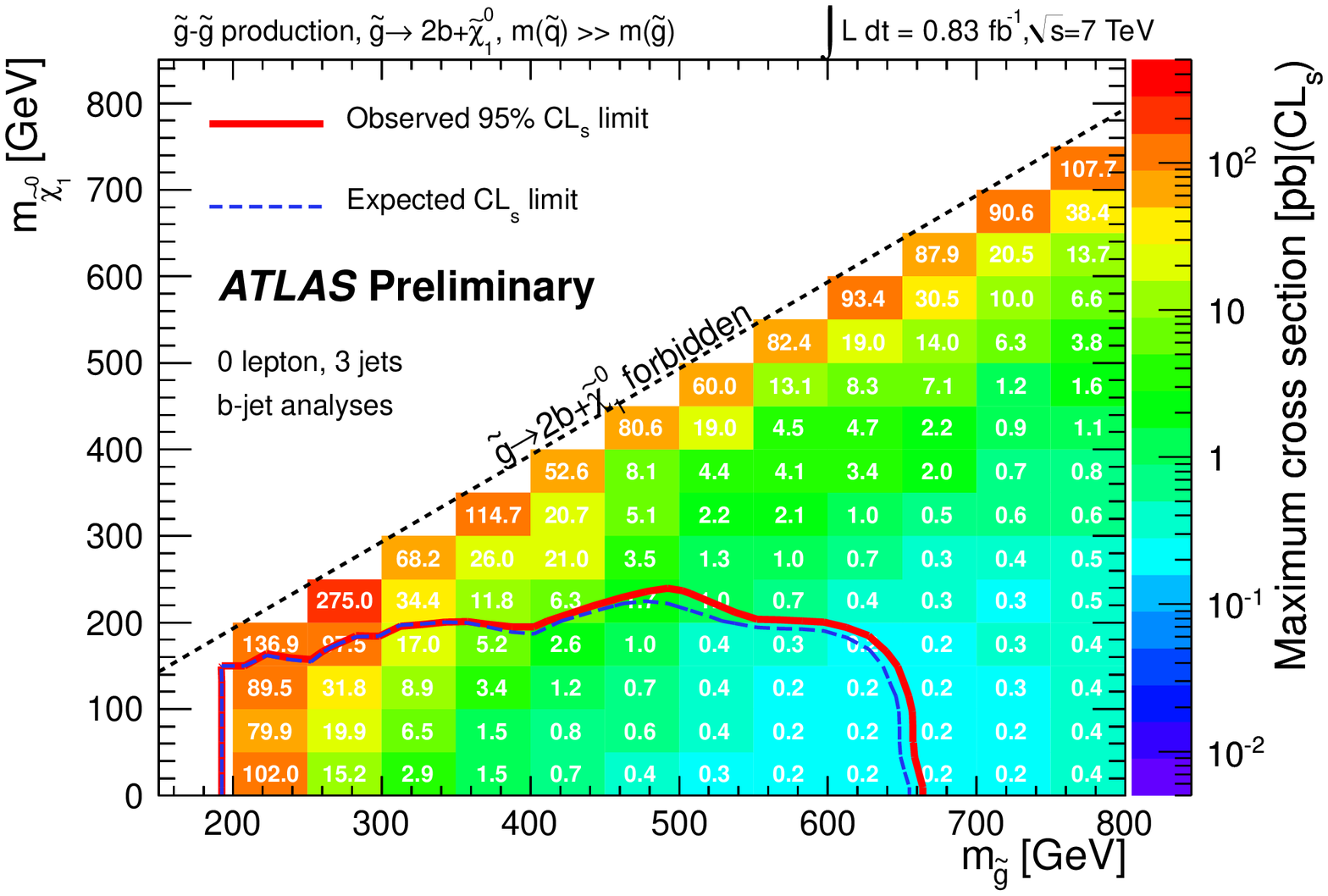}
\end{center}
\end{minipage}
\caption{Left: Limits from ATLAS \cite{ATLAS_bjet} in the
  gluino-sbottom mass plane, in a simplified model where $\tilde{g}
  \rightarrow \tilde{b}_{1}b$ and $\tilde{b}_{1} \rightarrow
  b\tilde{\chi}_{1}^{0}$. Right: Limits on the production cross section as a
  function of gluino and LSP mass in a model where all the squarks are
heavier than the gluino and gluino decays purely via the three-body
decay $\tilde{g} \rightarrow b\overline{b}\chi_{1}^{0}$ via an
off-shell sbottom.  The contours show the region
  excluded by the ATLAS analysis.}
\label{fig:bjet}
\end{figure}

ATLAS has updated the search in b-tagged jets plus $E_{T}^{miss}$,
asking for 3 or more jets, of which one is b-tagged \cite{ATLAS_bjet}.  Four
signal regions are defined,
based on the number of btags versus the scalar sum of $E_{T}^{miss}$
and the $p_{T}$ of the jets in the event.  The main background from
$t\overline{t}$ is estimated from Monte Carlo.  QCD multijet
background is evaluated in a similar way to the ATLAS search in the
jets+$E_{T}^{miss}$ channel, taking into account the differences in the jet
response function between light-quark and b-jets.
The null search is interpreted in a simplified SUSY model
containing gluinos, partners of the b quarks (sbottoms),
and the LSP, with everything else set to
high mass.  Assuming 100\% branching ratio of 
$\tilde{g} \rightarrow \tilde{b}_{1}b$ and $\tilde{b}_{1} \rightarrow
b + \tilde{\chi}_{1}^{0}$, limits are set in the plane of
gluino vs sbottom mass, assuming a $\tilde{\chi}_{1}^{0}$ mass of 60 GeV; the
results are shown in Fig. \ref{fig:bjet} (left). Gluino masses below
720 GeV are excluded for sbottom masses up to 600 GeV.  Fig. \ref{fig:bjet}
(right) shows the exclusion limits that
are placed in the gluino-LSP mass plane, as well as the limits on the
cross section in the same plane, in a model where all the squarks are
heavier than the gluino and gluino decays purely via the three-body
decay $\tilde{g} \rightarrow b\overline{b}\chi_{1}^{0}$ via an
off-shell sbottom.

\begin{figure}
\begin{minipage}{0.495\linewidth}
\begin{center}
\includegraphics[width=0.95\linewidth]{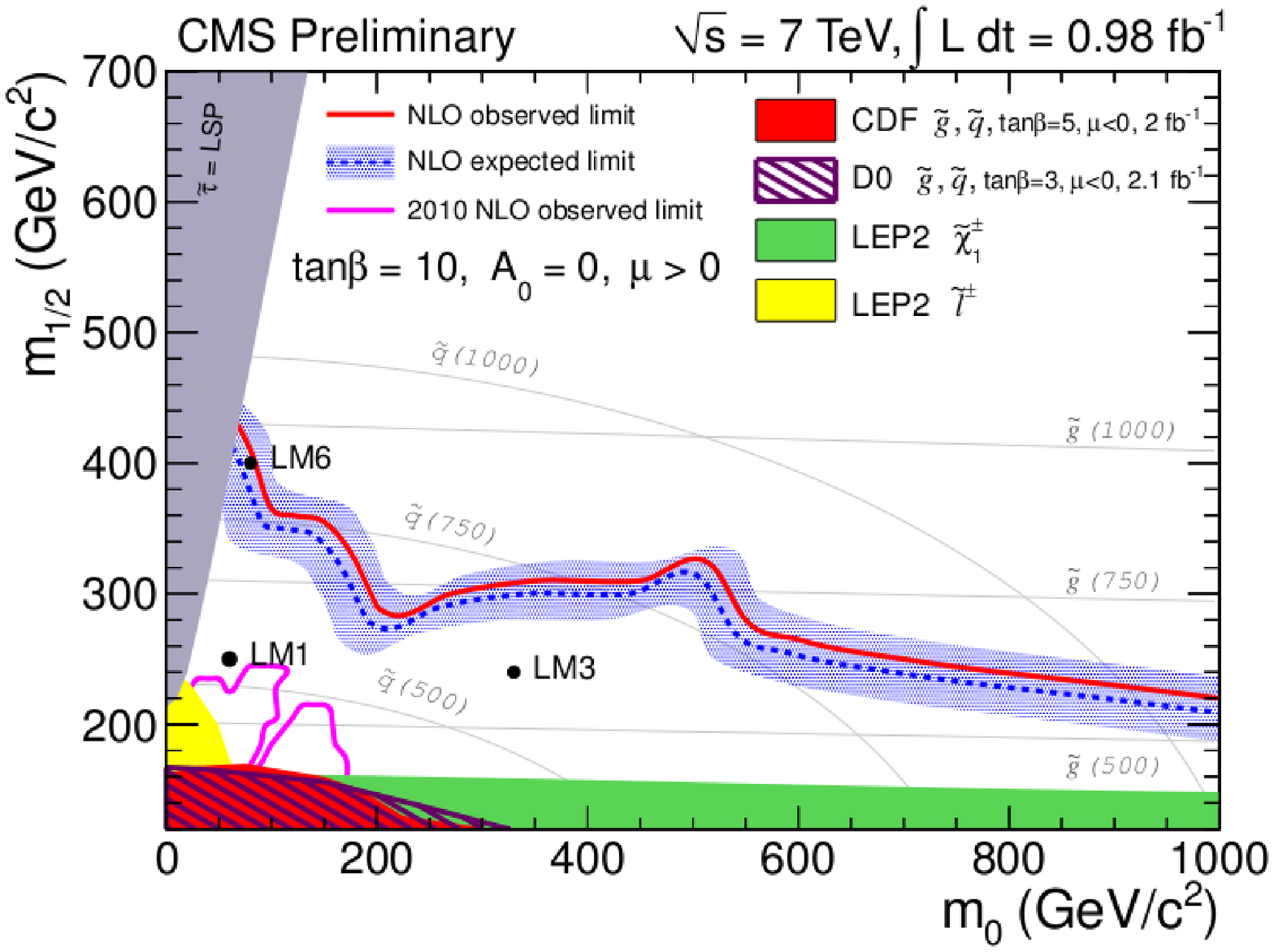}
\end{center}
\end{minipage}
\begin{minipage}{0.495\linewidth}
\begin{center}
\includegraphics[width=0.95\linewidth]{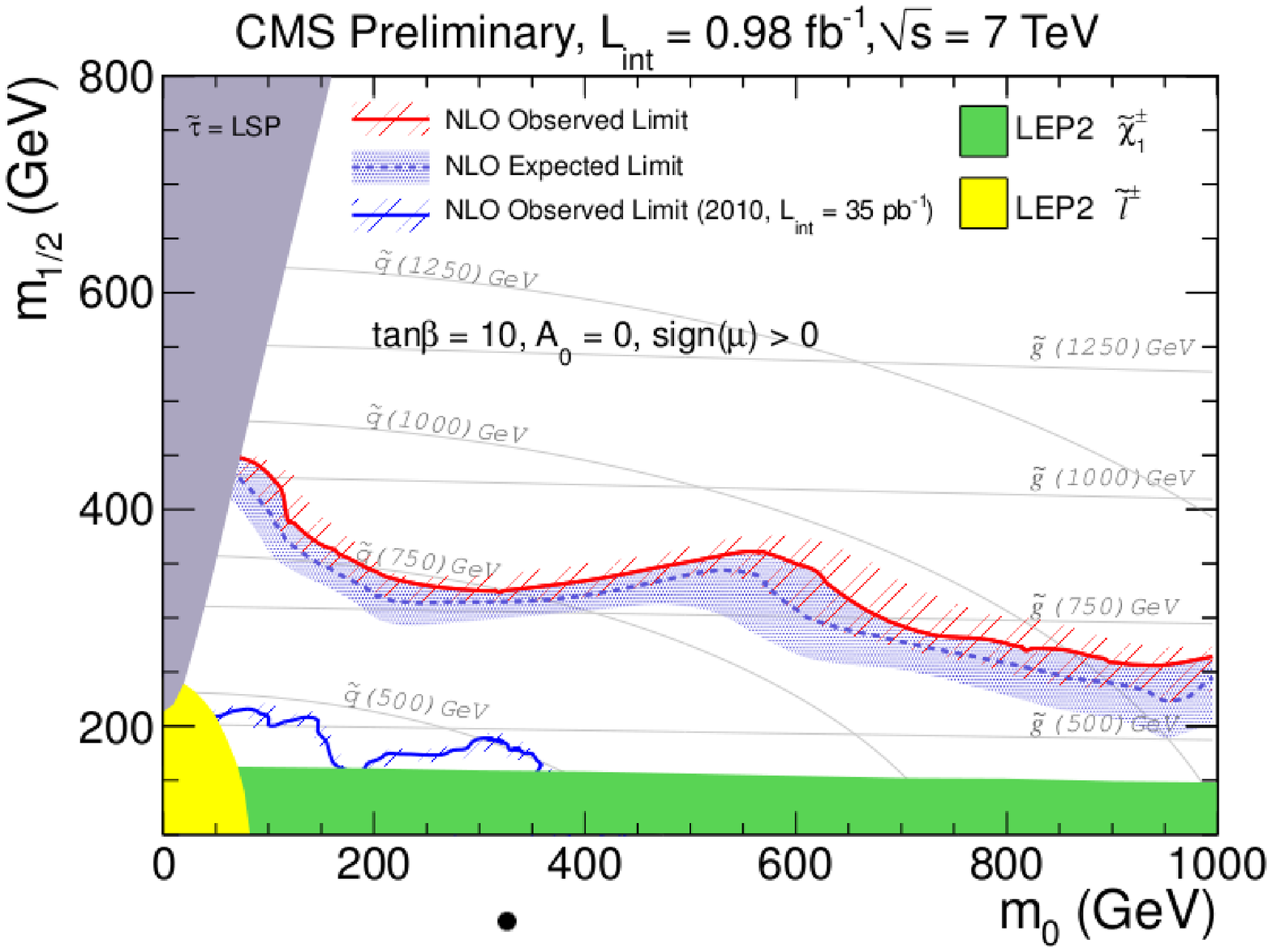}
\end{center}
\end{minipage}
\caption{Limits in the MSUGRA/CMSSM model from CMS in the OS dilepton
  (left) \cite{CMS_2lOS} and SS dilepton (right) \cite{CMS_2lSS}
  channels. }
\label{fig:2l}
\end{figure}

CMS has updated the results in the dilepton + jets + $E_{T}^{miss}$ channel, for
both opposite-sign (OS) \cite{CMS_2lOS} and same-sign (SS)
\cite{CMS_2lSS} dileptons; a new search in the $Z+E_{T}^{miss}$+jets
channel has also become available \cite{CMS_ZMET}.
In the OS search, two signal regions are defined,
one with $H_{T} > 300$ GeV and $E_{T}^{miss} > 275$ GeV and the
other with $E_{T}^{miss} > 200$ GeV and $H_{T} > 600$ GeV.
The main background from $t\overline{t}$ is estimated using an
ABCD method exploiting the lack of correlation between $H_{T}$ and
$E_{T}^{miss}$ 
significance.   The background estimates are cross-checked with two
other methods.  The first uses the observed $p_{T}$ spectrum
of the dilepton system to infer the $p_{T}$ of the two-neutrino system
from dileptonic $t\overline{t}$ decays.  The second looks at the
difference in the number of same-flavor versus opposite-flavor
dilepton events, suitably corrected for the relative efficiency for
detecting electrons versus muons.  The final background
estimate is
taken as the uncertainty-weighted average of the two methods described
above.  Limits are placed in the
MSUGRA/CMSSM model, and are shown in Fig. \ref{fig:2l} (left).

\begin{figure}
\begin{minipage}{0.495\linewidth}
\begin{center}
\includegraphics[width=0.95\linewidth]{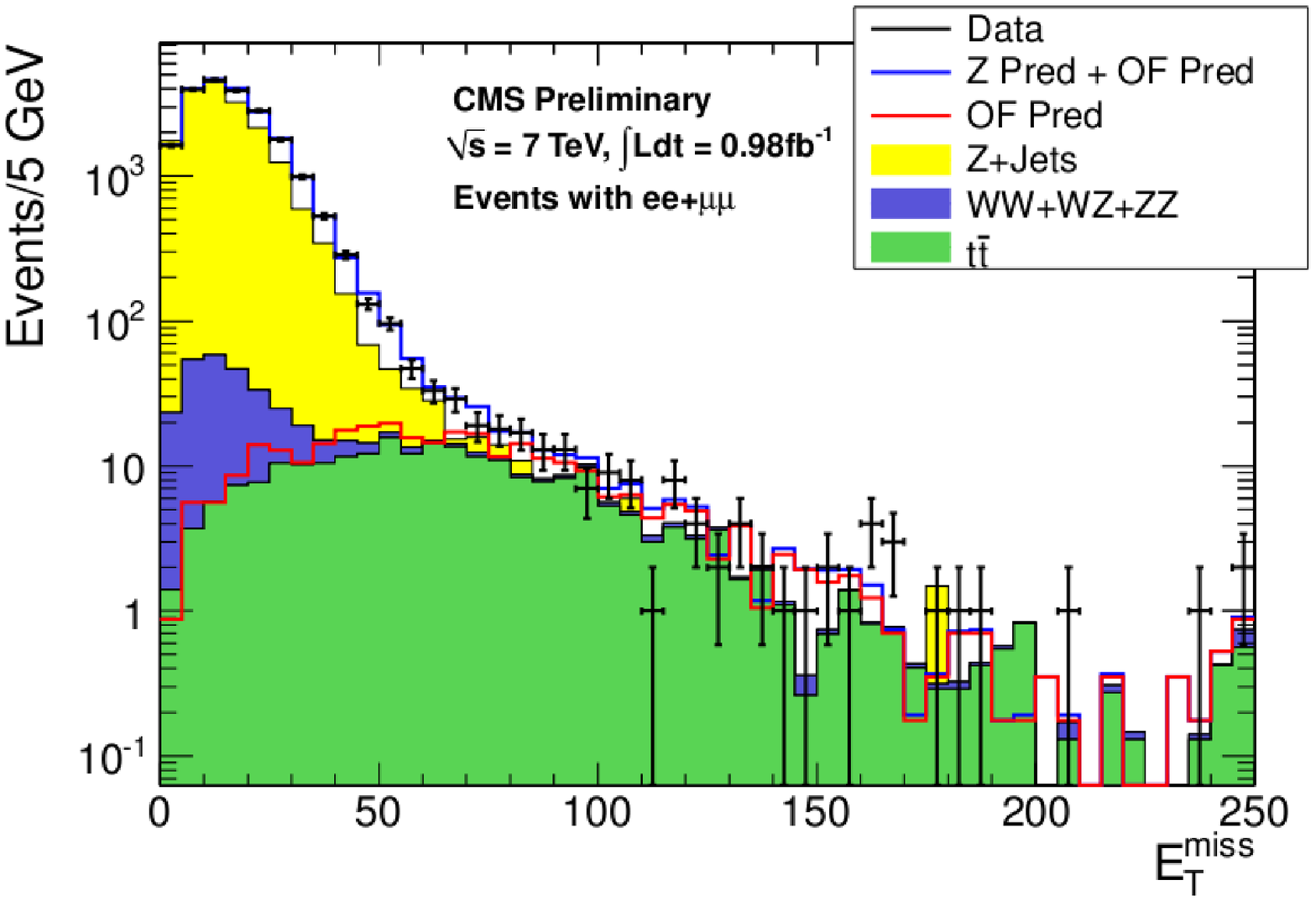}
\end{center}
\end{minipage}
\begin{minipage}{0.45\linewidth}
\begin{center}
\includegraphics[width=0.95\linewidth]{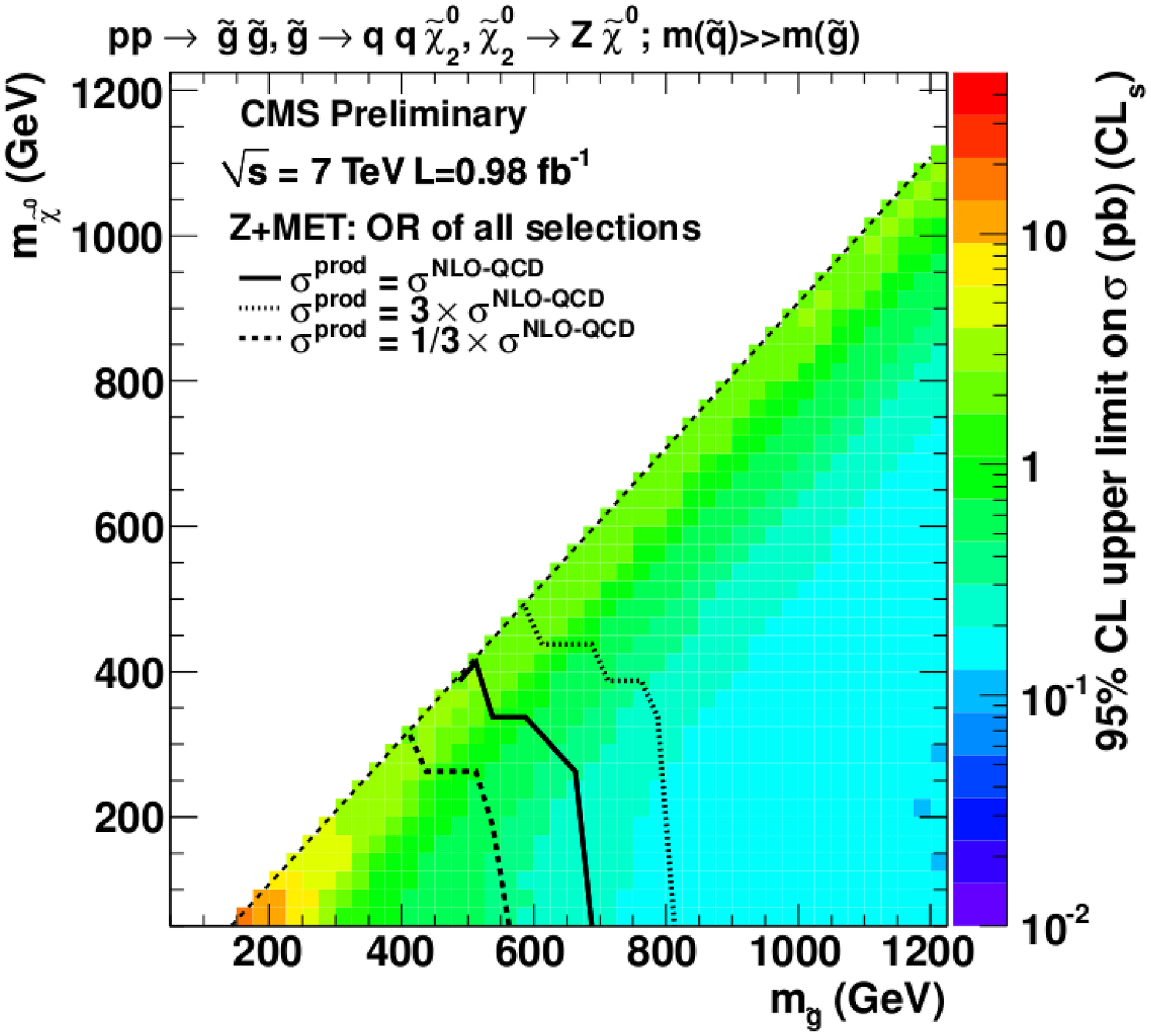}
\end{center}
\end{minipage}
\caption{Left: observed $E_{T}^{miss}$ distribution from the CMS
  $Z+E_{T}^{miss}$+jets analysis \cite{CMS_ZMET}.
  Right: Cross section limits in the gluino-LSP plane,
  assuming a simplified model of gluino pair
  production, followed by the decay $\tilde{g} \rightarrow
  qq\tilde{\chi}_{2}^{0}$ and $\tilde{\chi}_{2}^{0} \rightarrow
  Z\tilde{\chi}_{1}^{0}$.}
\label{fig:ZMET}
\end{figure}

The $Z+E_{T}^{miss}$+jets analysis requires a pair of same-flavor
leptons consistent with coming from a $Z$, at least two jets with
$p_{T} > 30$ GeV and $E_{T}^{miss} > $ 100 or 200 GeV.  The dominant
background comes from $t\overline{t}$ and is estimated from the yield
of opposite-flavor leptons, corrected for the relative efficiency for
detecting electrons versus muons.  The small background from $Z$+jets is
estimated using a control sample of $\gamma$+jets events.  The
observed $E_{T}^{miss}$ distribution is shown in Fig. \ref{fig:ZMET}
(left), compared to expectations.  Limits are
placed on a simplified model of gluino pair production with the decay
$\tilde{g} \rightarrow qq\tilde{\chi}_{2}^{0}$ and $\tilde{\chi}_{2}^{0} \rightarrow
Z\tilde{\chi}_{1}^{0}$; the limit contours are shown in Fig. \ref{fig:ZMET}
(right). Additional information on the detector response (lepton
reconstruction and isolation efficiencies, as well as the detector
response for $E_{T}^{miss}$) is provided so that Monte Carlo
generator-level samples can be compared  to the dilepton
observations.

In the same-sign dilepton search, CMS defines a number of signal regions, again
based mainly on $H_{T}$ and $E_{T}^{miss}$.  The background from fake leptons is
estimated using the standard matrix method.
Via the use of a hadronic trigger, regions of
low lepton $p_{T}$ can be probed.  Taus are also included in the
  analysis.    Limits are placed in the
MSUGRA/CMSSM model here as well, as shown in Fig. \ref{fig:2l}
(right).
In addition,
simple parametrizations of
the detector efficiency and resolution are provided,
allowing comparison of a variety
of Monte Carlo models with the experimental results.

\subsection{Outlook}

By the end of 2011, ATLAS and CMS are expected to have accumulated
approximately 4-5 $\rm{fb}^{-1}$ of data.  The 2012 run could extend
this to about 10 $\rm{fb}^{-1}$ per experiment.  BSM searches
will continue to push out to higher mass, but with steeply falling
cross sections, significant gains in mass reach will be hard to come
by.  More significant gains might be expected by pushing towards
smaller couplings, or in the case of SUSY, gaining access to new
production processes.  SUSY searches at the LHC have so far
concentrated on strong production of gluinos and
squarks of the first and second generations.  With higher integrated
luminosities, direct production of third generation squarks and direct
gaugino production should become accessible.
It should be noted that the Tevatron experiments
still have the best limits on stop \cite{CDF_stop, D0_stop}, sbottom
\cite{D0_sbottom}, and gaugino \cite{D0_gaugino} production.
Even direct
slepton production might be detectable.  Another priority is to extend
the SUSY searches to more challenging decay chains, those with small mass
differences in the decay cascade, or conversely highly boosted LSP's.

\section{Conclusion}

A very rich program of BSM searches continues at CDF and D0, with
typically 5-6 $\rm{fb}^{-1}$ analyzed to this point, out of 11.5
$\rm{fb}^{-1}$ delivered.  Best
limits on new physics are still coming from the Tevatron in a number
of cases.  The attention of the community has focused on a  few
recent anomalies from the Tevatron.
Whether these are ``mirages'' or signs of new physics
remains to be seen; the analyses are being followed up with the full
dataset.  Crosschecks by the LHC experiments will provide further
information. 

At the LHC, as of the end of July, ATLAS and CMS are starting to
produce results with 1 $\rm{fb}^{-1}$ of data analyzed.  Both
experiments are exploring a wide variety of signatures and
trying out new ways to present their results in as
model-independent manner  as possible.  There is unfortunately no sign
of new physics yet from the LHC.

Returning to the desert theme that opened this note,
one of the most famous desert stories in the Western canon is
the biblical story in Exodus in which
Moses finally catches a glimpse of
the Promised Land after 40 years of
wandering in the desert.  In this context, perhaps it is worth recalling
that this year marks the 40th year since the birth of SUSY \cite{SUSY}
and the 37th anniversary\footnote{To first order, 40 years.} of
the ``November revolution'' \cite{Ting,
  Richter}.  Long ago, the wise ones in our field promised
  the ``LHC no lose theorem''.  It can only be hoped that this is the
  year in which the Promise is realized.\footnote{But if we insist on
    40 years since the ``November revolution'' this would imply
    having to wait for 14 TeV running at the LHC!}

\bigskip 

\end{document}